\documentclass[11pt]{revtex4}
\usepackage{amsmath}
\usepackage{amsfonts}
\usepackage{amssymb}
\begin{document}
\title{Massless and massive representations in the {\it spinor technique}
}
\author{T. Troha, D. Lukman and N.S. Manko\v c Bor\v stnik\\
University of Ljubljana,\\ Slovenia 
}

\begin{abstract} 
The technique~\cite{norma92,norma93,hno2,hno6,norma94,norma95,pikanorma} for representing spinors 
and the definition of the discrete 
symmetries~\cite{HNds} is used to illustrate on a toy model~\cite{DHN,DN012,HN,hn07} properties 
of massless and massive spinors states, in the first and the second quantized picture. 
Since in this toy model the number of the starting massless representations is well 
defined as well as the origin of masses and charges in $d=(3+1)$ space, this contribution 
might help to clarify the problem about Dirac, Weyl and Majorana 
kinds of representations~\cite{Barut,dvoeglazov,DVMajorana,Ni} in physically more interesting 
cases%, which we
%also comment by using the {\it spin-charge-family} theory~\cite{norma}
. 
\end{abstract}
%\maketitle
%

\keywords{Spinor representations, Kaluza-Klein theories, Discrete symmetries, Higher dimensional spaces,
Beyond the standard model}

\pacs{11.30.Er,11.10.Kk,12.60.-i, 04.50.-h
%%
%12.15.Ff   12.60.-i  12.90.+b   11.30.Hv  12.15.-y  11.30.-j  14.80.-j
}

\maketitle
\section{Introduction}
\label{introduction} 

We illustrate on a toy model, defined in  %(an even dimensional space-time) $d=(5+1)$, presented in 
the refs.~\cite{DHN,DN012,HN,hn07}, massless and massive,  positive and negative energy solutions 
of the equations of motion for spinors (fermions), and study the properties of particles and the 
corresponding antiparticles states, by taking into account the definition of 
the discrete symmetry operators (${\underline {\bf \mathbb{C}}}_{{\bf {\cal N}}}$,
${\cal P}_{{\cal N}}$ and ${\cal T}_{{\cal N}}$) in the second quantized picture, designed in the 
paper~\cite{HNds} for higher dimensional spaces. For the second quantized picture we use, like in 
the paper~\cite{HNds}, the concept of the Dirac sea, since it offers a nice physical understanding.  

We use this toy model, in which we start with one spinor Weyl representation and assume the 
${\cal M}^{5+1}$ manifold to break into ${\cal M}^{3+1}$ $\times$ an almost $S^2$ sphere due to 
the zweibein and spin connection fields in $d=(5,6)$, since  in this toy model the origin of 
the charge is the spin or the total angular momentum in extra $d=(3+1)$ dimensions ($d>(3+1)$) 
and the mass originates either in the dynamics in the higher dimensions or in the vacuum 
expectation values of the scalar fields, which are the gauge fields with the scalar index with 
respect to $d=(3+1)$, and therefore well defined, making discussions about the representations 
of spinors well defined and transparent. 
The illustration makes evident the fact that without knowing the action which leads to massless 
and massive solutions, the analyse of the Dirac equation which already assumes the mass and do not 
pay attention on the origin of the charge might not help a lot.

We comment also the representations in $d=(7+1)$ and $d=(13+1)$.  In this last case one Weyl 
representation of massless states contains left handed weak charged quarks and leptons as well 
as right handed weak chargeless quarks and leptons, that is all the family members postulated 
by the {\it standard model}, with right handed neutrinos included, while the simple starting 
Lagrange density for spinors and gauge fields offers the explanation for the {\it standard 
model} as the low energy effective theory of this {\it spin-charge-family} theory of one of us 
(S.N.M.B.)~\cite{NBled2013,norma,norma94,norma95,pikanorma,gmdn07,gn}, with the families included.

Assuming the Lorentz invariance and causality, the theorem of CPT is generally valid, while the 
$C$, $P$ and $T$ symmetries, separately or in pairs, depend on effective theories.
%and so do other discrete symmetries. 
In our  toy model~\cite{DHN,DN012,HN,hn07}, like in all Kaluza-Klein kind of models 
in even dimensional spaces (in odd dimensional spaces there is no mass protection mechanism 
and they do not lead  
%after the break of symmetries, 
effectively to observable phenomena, that is to almost massless quarks and leptons),
besides ${\underline {\bf \mathbb{C}}}_{{\bf {\cal N}}}\times$ ${\cal P}_{{\cal N}}\times$ 
${\cal T}_{{\cal N}}$, only ${\underline {\bf \mathbb{C}}}_{{\bf {\cal N}}}\times$ ${\cal P}_{{\cal N}}$ 
and  ${\cal T}_{{\cal N}}$ are good symmetries, since only these symmetries operate among the 
eigenstates of the equations of motion belonging to the starting spinor representation. 
The Clifford odd operators, like $\gamma^0, \gamma^0 \gamma^5$ (we use $\Gamma^{(3+1)}$ instead 
of $\gamma^5$) in the refs.~\cite{Barut,dvoeglazov,DVMajorana,Ni}, namely transform the solutions 
into another Weyl representation, which only makes sense if the equations of motion contain the 
operators which connect both representations. 

The representations in the 
{\it spinor technique}~\cite{norma92,norma93,hno2,hno6,norma94,norma95,pikanorma,norma} makes our 
illustration of free as well as interacting particles and antiparticles, massless and massive in 
$d=(3+1)$, with  their discrete symmetries included, easier to follow. 

We do not study in this paper the families of spinors. Families, their number is  in even dimensional 
spaces equal to $2^{\frac{d}{2}-1}$ (and therefore in $d=(5+1)$ equal to $4$ in our toy model) 
would not clarify much the point of this paper, since they form equivalent representations with 
respect to the here presented states~\footnote{The {\it spin-charge-family} 
theory~\cite{norma,NBled2013,norma92,norma93,norma94,norma95,pikanorma,gmdn07,gn,hno3} of one of us 
(S.N.M.B.) offers the explanation not only for the appearing of charges and gauge vector and scalar 
fields but also for the appearance of families.}.   
%This contribution is the improved version of the ref.~\cite{TDN}. 
%
One Weyl representation in an even dimensional space  contains $2^{\frac{d}{2}-1}$ states, the same number 
as there is the  number of families, which is in the case of our toy model in  $d=(5+1)$,  with 
${\cal M}^{3+1}$ $\times$ an almost $S^2$ sphere, 
equal to $4$.

Following the refs.~\cite{HNds,DHN,DN012,HN,hn07} we present in sect.~\ref{toy} the action for a 
spinor (fermion) in 
$d=(5+1)$, leading to the Weyl equations which manifest  mass protected massless with the spin in 
$(5,6)$ dimension as the charge in $d=(3+1)$ state and the series of states with the total angular 
moments in $d=(5,6)$ as the charges~(\ref{chargemassless}). 
%We discuss properties of these massive and massless states with the conserved Kaluza-Klein charges. 
We look also for the massive chargeless solutions of the Weyl equations, for masses of which are 
responsible, by the assumption, nonzero vacuum expectation values of the scalar 
fields~\cite{norma,NBled2013,hno3,gmdn07} - spin connections and zweibeins with the scalar index with 
respect to $d=(3+1)$ (subsect.~\ref{chargelessmassive}), which are the gauge fields of $S^{56}$, 
playing the role of the higgs in the {\it standard model} and carrying in this toy model case only
the hyper charge $Y$, determined by the operator $S^{56}$. The particle, antiparticle and Majorana 
states~(\ref{Majorana}) are presented. 

In subsect.~\ref{discrete} the discrete symmetry operators defined in the ref.~\cite{HNds} are presented.

In subsect.~\ref{Dirac} the representations are commented from the point of view of the usual Dirac 
ones~\cite{ItZu}.

We conclude our paper with discussions (sect.~\ref{conclusions}) on the (starting) Weyl representation
of the toy model and the corresponding the first and the second quantized states as they manifest in 
$d=(3+1)$. We extend  discussions also on the cases in $d=(7+1)$ and $d=(13 + 1)$ and comment in 
all these cases the action of the Clifford odd operators on the states.

\section{The toy model with the manifold $M^{5+1}$ broken into $M^{3+1}\times$ an almost $S^2$ 
and the representations} 
\label{toy}

We make a choice of the action for massless (Weyl) spinors~\cite{DHN} living on the manifold $M^{1+5}$ 
\begin{eqnarray}
\label{action}
S&=& \int d^{d}x \,  {\cal L}_{W}\,, \nonumber\\
{\cal L}_{W}&=& \psi^{\dagger}\,E\, \gamma^0 \gamma^a  \{f^{\alpha}{}_a p_{\alpha} +
\frac{1}{2E} \{p_{\alpha},f^{\alpha}{}_a E\}_-   -\frac{1}{2} S^{cd}  \omega_{cda} 
 \}\psi\,\nonumber\\
&=& \frac{1}{2}  \, \Psi^{\dagger} \,E\, \gamma^0 \, (\gamma^m \, p_{0m} 
+ \gamma^s \, p_{0s})\,\,\Psi \,+ h.c.\,,
%\quad  +\; {\rm the} \; {\rm rest},
\nonumber\\
p_{0m}   &=&  p_{m} - S^{56} \,\omega_{56 m}\,,\nonumber\\
p_{0s}   &=&  f^{\sigma}_{s} \,p_{\sigma} + \frac{1}{2E} \{p_{\sigma},f^{\sigma}_{s} E \}_{-}
-  S^{56} \,\omega_{56 s}\,,\nonumber\\
m &=& (0,1,2,3)\,, \quad s=(5,6)\,.
\end{eqnarray}
$f^{\alpha}{}_a $ are vielbeins and $\omega_{cda} $ spin connection fields, the gauge fields 
of the moments $p_{\alpha}$ and $S^{ab}$, respectively.  We take flat $(3+1)$ space: $f^{\mu}{}_m 
=\delta^{\mu}_{m}\,$, $\, \mu= ((0),(1),(2),(3))$,  $m=(0,1,2,3)$. 
In our toy model the manifold $M^{5+1}$ breaks into $M^{3+1} \times $ (an almost) $S^2$. 
$S^{56}$ then manifests in $d=(3+1)$ as the (Kaluza-Klein) charge, $\omega_{56m}$ as the 
corresponding vector and $\omega_{56 s}$ as the corresponding scalar gauge fields. Not paying 
attention to the family quantum numbers, we left in Eq.~(\ref{action}) out all the terms which 
carry family quantum numbers~\footnote{The {\it spin-charge-family} theory assumes the presence
of the  fields $\tilde{\omega}_{abc}$, the gauge fields of the generators $\tilde{S}^{ab}= $ 
$\frac{i}{4} (\tilde{\gamma}^a \tilde{\gamma}^b)-\tilde{\gamma}^b \tilde{\gamma}^a)$, with 
$\{\tilde{\gamma}^a,\gamma^b)\}_{-}=0$. This second kind of the Clifford algebra 
objects~\cite{norma92,norma93,hno3,hno6} form obviously the equivalent representations 
with respect to the Dirac one. There are only two kinds of gamma operators: the Dirac one 
corresponding to the left multiplication of any Clifford algebra object and this second one 
corresponding to the right multiplication of the same Clifford algebra object.}  and determine  
correspondingly interaction among different families.    
% norma93,dirac-k,hn0203,dnBledfamilies

The Weyl equation of the action~(Eq.(\ref{action})) can be %in the case of our toy model 
written as follows
 \begin{eqnarray}
 \label{weyl}
 &&(\gamma^m p_{m} + \stackrel{56}{(+)}p_{0+} + \stackrel{56}{(-)} p_{0-}) \psi=0\,,\nonumber\\
 &&p_{0\pm}= p_{0}^5 \mp i\,p_{0}^6\,,\nonumber\\
 &&p_{0s}= f^{\sigma}_{s}(p_{\sigma} -\frac{1}{2} S^{ab} \omega_{ab \sigma}) + 
 \frac{1}{2E} \{p_{\sigma},f^{\sigma}_{s} E \}_{-} \,,\nonumber\\
 &&\stackrel{56}{(\pm)} = \frac{1}{2}\,(\gamma^5 \pm i \gamma^6)\,.
 %- \frac{1}{2}  \tilde{S}^{ab}\tilde{\omega}_{ab \sigma})
 \end{eqnarray}
The explanation how does the {\it technique}~\cite{norma92,norma93,hno2,hno6,norma94,norma95,pikanorma} 
work can be found in 
the refs.\cite{norma92,norma93,hno2,hno6,norma94,norma95,pikanorma}, 
the short version, taken from the ref.~\cite{norma,NBled2013,gmdn07,HNds}, is in the appendix~\ref{technique}. 

%We do not pay attention in this paper on the family degrees of freedom.

There are $2^{\frac{d}{2}-1}$ ($4$ in our case of $d=6 $) basic spinor states~\footnote{According to 
the {\it spin-charge-family} theory there is the same number of families as there are spinor states 
within one family representation: $2^{\frac{d}{2}-1}$.}, forming the fundamental representation 
of the group $SO(5,1)$, the generators of which are $S^{ab}= \frac{i}{4}(\gamma^a \gamma^b -
\gamma^b \gamma^a)$
\begin{eqnarray}
\Psi_{1} &=&  \stackrel{03}{(+i)} \stackrel{12}{(+)} \stackrel{56}{(+)}|vac>_{fam},\nonumber\\
\Psi_{2} &=&\stackrel{03}{(+i)} \stackrel{12}{[-]} \stackrel{56}{[-]}  |vac>_{fam},\nonumber\\
\Psi_{3} &=&  \stackrel{03}{[-i]} \stackrel{12}{[-]} \stackrel{56}{(+)} |vac>_{fam},\nonumber\\
\Psi_{4} &=& \stackrel{03}{[-i]} \stackrel{12}{(+)}\stackrel{56}{[-]}|vac>_{fam}\,,
\label{weylrep}
\end{eqnarray}
with $|vac>_{fam}$ defined so that these four states are nonzero and normalized:
$\Psi^{\dagger}_{i} \Psi_{j}= \delta^{i}_{j}$. 
(This vacuum is not a second quantized vacuum.) 
All the basic states are eigen states of the Cartan subalgebra (of the Lorentz transformation 
Lie algebra), for which we take: $S^{03}, S^{12}, S^{56}$, with the eigen values, which can be 
read from Eq.~(\ref{weylrep}) if taking $\frac{1}{2}$ times the numbers $\pm i$ or $\pm 1$ 
in the parentheses of nilpotents $\stackrel{ab}{(k)} $ and projectors $\stackrel{ab}{[k]} $: 
$S^{ab} \stackrel{ab}{(k)} =$ $\frac{k}{2} \stackrel{ab}{(k)} $, $S^{ab} \stackrel{ab}{[k]} =$ 
$\frac{k}{2} \stackrel{ab}{[k]}$. One notices that two are right $(\Psi_{1} $ and $\Psi_{3}) $
and two left $(\Psi_{2} $ and $\Psi_{4}) $ handed with respect to $d=(3+1)$ (while all four 
carry the same, left, handedness with respect to $d=(5+1)$). The operator of handedness is 
defined in Eq.~(\ref{handedness}). The operator with an odd number of the 
Clifford algebra objects, like  it is $\gamma^0 \Gamma^{(3+1)}$ ($\gamma^0 \gamma^{5}$ in usual notation) 
would transform any state into the state of the opposite handedness.

The following choice of the zweibein fields causes that the infinite surface $d=(5,6)$ curls 
into an almost $S^2$ (with one hole~\cite{DHN})
\begin{eqnarray}
e^{s}{}_{\sigma} &=& f^{-1}
\begin{pmatrix}1  & 0 \\
 0 & 1 
 \end{pmatrix},
f^{\sigma}{}_{s} = f
\begin{pmatrix}1 & 0 \\
0 & 1 \\
\end{pmatrix}\,,
\label{fzwei}
f = 1+ (\frac{\rho}{2 \rho_0})^2\,,\nonumber\\ 
E &=& \det(e^s{\!}_{\sigma})=f^{-2}\,, e^s{\!}_{\sigma}\,f^{\sigma}{\!}_{t}= \delta^{s}_{t}\,,
%= \frac{2}{1+\cos \vartheta}\,,
\nonumber\\ 
x^{(5)} &=& \rho \,\cos \phi,\quad  x^{(6)} = \rho \,\sin \phi\,, \nonumber
\end{eqnarray}
while $d=(3+1)$ space stays flat ($f^{\mu}{\!}_{m}= \delta^{\mu}_{m}$). We choose the 
spin connection fields on this $S^2$ as 
\begin{eqnarray}
  f^{\sigma}{}_{s'}\, \omega_{st \sigma} &=& i F\, f \, \varepsilon_{st}\; 
  \frac{e_{s' \sigma} x^{\sigma}}{(\rho_0)^2}\, , \quad 
 0 <2F \le 1\, 
  %= -i \varepsilon_{st}\, \frac{f \,F\,\sin \vartheta}{\rho_0}\, (\cos \phi,\sin \phi)
  ,\quad s=5,6,\,\,\; \sigma=(5),(6)\,, 
\label{omegas}
\end{eqnarray}
in order to guarantee that there manifests in $d=(3+1)$ only one massless and correspondingly mass 
protected state~\cite{DHN}, while the rest of states are all massive. There is the whole interval 
for the constant $F$ ($0 <2F \le 1$), which fulfills the condition of only one massless state of 
the right handedness in $d=(3+1)$, which is square integrable. 
%To simplify our discussions we shall make a choice in this paper of $F= \frac{1}{2}$.

When requiring that the solutions of Eq.~(\ref{weyl}) have the angular moments in $d=(5,6)$ 
manifesting the charges in $d=(3+1)$ ($M^{56}= x^5 p^{6}- x^6 p^{5}+ S^{56}=$ 
$- i\frac{\partial}{\partial \phi} + S^{56}$), we write the wave functions  $\psi^{(6)}_{n+1/2}$ 
for the choice of the coordinate system $p^a= (p^0,0,0,p^3,p^5,p^6)$ as follows 
\begin{eqnarray}
\psi^{(6) }_{n+1/2}= ({\cal A}_{n}\,\stackrel{03}{(+i)} \stackrel{12}{(+)} \stackrel{56}{(+)}  
+ {\cal B}_{n+1}\, e^{i \phi}\, \stackrel{03}{[-i]}\stackrel{12}{(+)}\,\stackrel{56}{[-]})\,\cdot e^{in \phi}
e^{-i(p^0 x^0- p^3 x^3)}\,. 
\label{mabpsi}
\end{eqnarray}
Besides one massless ($\psi^{(6) }_{1/2}$) there is the whole series of massive solutions manifesting 
in $d=(3+1)$ the (Kaluza-Klein) charge ${n+1/2}$: $M^{56}\, 
\psi^{(6) }_{n+1/2}=(n+1/2)\,\psi^{(6) }_{n+1/2} $, and solve  Eq.~(\ref{weyl}), provided that  
${\cal A}_{n}$ and ${\cal B}_{n+1}$ are the solutions of the equations
\begin{eqnarray}
&&-if \,\{ \,(\frac{\partial}{\partial \rho} + \frac{n+1}{\rho})  -   
  \frac{1}{2\, f} \, \frac{\partial f}{\partial \rho}\, (1+ 2F)\}  {\cal B}_{n+1} + m {\cal A}_n = 0\,,  
\nonumber\\
&&-if \,\{ \,(\frac{\partial}{\partial \rho} - \quad \frac{n}{\rho}) -   
  \frac{1}{2\, f} \, \frac{\partial f}{\partial \rho}\, (1- 2F)\}  {\cal A}_{n} + m {\cal B}_{n+1} = 0\,.
\label{equationm56}
\end{eqnarray}
The  massless positive energy solution with spin $\frac{1}{2}$, left handed (Eq.~(\ref{handedness}) in 
$d=(5+1)$, the charge in $d=(3+1)$ equal to $\frac{1}{2}$ and right handed with respect to 
$\Gamma^{(3+1)}$ is equal to
\begin{eqnarray}
\psi^{(6)}_{\frac{1}{2}} ={\cal N}_0  \; f^{-F+1/2} 
\stackrel{03}{(+i)}\stackrel{12}{(+)}\stackrel{56}{(+)}\,e^{-i(p^0 x^0-p^3x^3)}\,. 
\label{Massless}
\end{eqnarray} 
For the special choice of $F=\frac{1}{2}$ (from the interval in Eq.~(\ref{omegas}) allowing  
only right handed square integrable massless states) the solution (Eq.~(\ref{Massless})) simplifies 
to
\begin{equation}
\label{mlessF}
\psi^{(6)}_{\frac{1}{2}} ={\cal N}_0  \;  
\stackrel{03}{(+i)}\stackrel{12}{(+)}\stackrel{56}{(+)}\,e^{-i(p^0 x^0-p^3x^3)}\,.
\end{equation}

Massive solutions are in this special case~\cite{DHN,DN012} expressible in terms of the associate 
Legendre function $P^{l}_n(x)$, $x= \frac{1-u^2}{1+u^2}$, $u= \frac{\rho}{2 \rho_0}$, where $\rho_0$ 
is the radius of (an almost) $S^2$, as follows  
\begin{eqnarray}
\label{massF}
{\cal A}^{l(l+1)}_n &=& P^{l}_n\,,\nonumber\\
{\cal B}^{l(l+1)}_{n+1} &=& \frac{-i}{\rho_0 m} \,\sqrt{1-x^2}\, 
\left(\frac{d}{dx} + \frac{n}{1-x^2} \right)\, {\cal A}^{l(l+1)}_n \,,
\end{eqnarray}
with the masses~\footnote{In the case that $d=(5,6)$ is a compact $S^2$ sphere these massive 
solutions would make infinite spectrum with quantum numbers $(l,n)$, $l$ defining in $d=(3+1)$ 
the mass and $n+ \frac{1}{2}$ the Kaluza-Klein charge. In the case of an almost $S^2$ the spectrum 
start to stop when the energy approaches the strengths of the source which causes the vielbein 
leading to an almost $S^2$.} determined by $(\rho_0\, m)^2 =l(l+1)$ and $l=1,2,3,\dots$, $0 \le n < l$. 

We shall comment all the solutions, massless and massive, in the subsections of this section.

{\it Let us summarize this section and its subsections}: Starting with one Weyl massless representation 
and the action for a massless 
spinor in $d=(5+1)$ , we end up with one massless and a series of massive solutions in  $d=(3+1)$). 
All the solutions are, from the point of view of $d=(3+1)$ distinguishable according to their 
charges and masses and also $(p^0,p^1,p^2,p^3)$ and all belong to the  starting left handed
spinor representation of Eq.~(\ref{weylrep}). (The operator of handedness is presented in 
Eq.~(\ref{handedness}).)

We check~(\ref{discrete}) the discrete symmetries of the action~(Eq.(\ref{action})) and of the 
Weyl equation~(Eq.(\ref{weyl})), using the operators expressible with an even number of the 
Clifford algebra objects (even number of $\gamma^a$'s), so that the transformed states stay 
within the starting Weyl representation. 
Any Clifford odd operator 
%, like it is $\gamma^0 \Gamma^(3+1)$ ($\gamma^0 \gamma^(5)$ in usual notation) 
would, namely, transform any  state into a state of the opposite handedness in $d=(5+1)$, 
which would have meaning if there would be terms in the starting action (Eq.(\ref{action})) 
connecting states of different handedness, which is not the case.
%Such a term would require a  different kind of action or the same kind of action in a larger $d$.

From massless (subsect.~\ref{chargemassless}) or massive (subsect.~\ref{chargemassive}) positive 
energy solutions the negative ones  follow by the application of 
${\cal C}_{{ \cal N}}\cdot {\cal P}^{(d-1)}_{{\cal N}}$. The positive energy states, put on the 
top of the Dirac sea, represent particles. The antiparticle states follow from the particle ones 
either by emptying the negative energy state in the Dirac sea or directly from the particle state 
by the application of the operator  $\mathbb{C}_{{ \cal N}}\cdot {\cal P}^{(d-1)}_{{\cal N}}$  
and putting the obtained state on the top of the Dirac sea. Any antiparticle, massless or massive,
carries the opposite charge than the corresponding particle.

Assuming that the scalar fields gain nonzero vacuum expectation values, the massless solutions 
no longer  exist. The effective equations of motion~(Eqs.(\ref{Weylmvac}, \ref{Weylmvacsimple})) 
lead to two positive  energy states,
%representing particles if put on the top of the Dirac sea 
and two corresponding negative energy states, the holes in which represent antiparticle states, 
which are indistinguishable from the particle states and are indeed the Majorana particles. 
Not paying attention to the origin of the mass term any longer~(Eq.(\ref{Weylmvac})) (one of) 
the discrete symmetry operators have to be redefined.

%%%%

%
\subsection{Discrete symmetry operators} 
\label{discrete}

To discuss representations of particle and antiparticle states we must define the discrete 
symmetry operators in the second quantized picture. 
The ref.~\cite{HNds} proposes the definition of the discrete symmetries operators for the 
Kaluza-Klein kind of theories, for the first and the second quantized picture, so that the 
total angular moments in higher dimensions manifest as charges in $d=(3+1)$.
We shall use, as in the ref.~\cite{HNds}, the Dirac sea second quantized picture to make  
discussions transparent.

The ref.~\cite{HNds} proposes the following discrete symmetry operators
\begin{eqnarray}
\label{CPTN}
{\cal C}_{{\cal N}}  &= & \prod_{\Im \gamma^m, m=0}^{3} \gamma^m\,\, \,\Gamma^{(3+1)} \,
K \,I_{x^6,x^8,\dots,x^{d}}  \,,\nonumber\\
{\cal T}_{{\cal N}}  &= & \prod_{\Re \gamma^m, m=1}^{3} \gamma^m \,\,\,\Gamma^{(3+1)}\,K \,
I_{x^0}\,I_{x^5,x^7,\dots,x^{d-1}}\,,\nonumber\\
{\cal P}^{(d-1)}_{{\cal N}}  &= & \gamma^0\,\Gamma^{(3+1)}\, \Gamma^{(d)}\, I_{\vec{x}_{3}}
\,.
\end{eqnarray}
The operator of handedness in even $d$ dimensional spaces is defined as
\begin{eqnarray}
\label{handedness}
\Gamma^{(d)} :=(i)^{d/2}\; \prod_a \:(\sqrt{\eta^{aa}}\, \gamma^a)\,,
\end{eqnarray}
 with products of 
$\gamma^a$ in ascending order. We choose $\gamma^0$, $\gamma^1$ real, $\gamma^2$ imaginary, 
$\gamma^3$ real, $\gamma^5$ imaginary, $\gamma^6$ real, alternating imaginary and real up to 
$\gamma^d$ real. 
Operators $I$ operate as follows: $\quad I_{x^0} x^0 = -x^0\,$; $
I_{x} x^a =- x^a\,$; $  I_{x^0} x^a = (-x^0,\vec{x})\,$; $ I_{\vec{x}} \vec{x} = -\vec{x}\,$; $
I_{\vec{x}_{3}} x^a = (x^0, -x^1,-x^2,-x^3,x^5, x^6,\dots, x^d)\,$; 
$I_{x^5,x^7,\dots,x^{d-1}}$ $(x^0,x^1,x^2,x^3,x^5,x^6,x^7,x^8,
\dots,x^{d-1},x^d)$ $=(x^0,x^1,x^2,x^3,-x^5,x^6,-x^7,\dots,-x^{d-1},x^d)$; $I_{x^6,x^8,\dots,x^d}$ 
$(x^0,x^1,x^2,x^3,x^5,x^6,x^7,x^8,\dots,x^{d-1},x^d)$
$=(x^0,x^1,x^2,x^3,x^5,-x^6,x^7,-x^8,\dots,x^{d-1},-x^d)$, $d=2n$. 

${\cal C}_{{\cal N}}$ transforms the state, put on the top of the Dirac sea, into the corresponding 
negative energy state in the Dirac sea.

We need the operator, we name~\cite{NBled2013,TDN,HNds} it $\mathbb{C}_{{ \cal N}}$, which transforms 
the starting single particle state on the top of the Dirac sea into the negative energy state and 
then empties this negative energy state.  
This hole in the Dirac sea is the antiparticle state  
put on the top of the Dirac sea. Both, a particle and its antiparticle state (both put on the top of 
the Dirac sea), must solve the Weyl equations of motion.

This $\mathbb{C}_{{ \cal N}}$ is defined as a product of the operator~\cite{NBled2013,TDN} $"emptying"$,
(which is really an useful operator, although it is somewhat difficult to imagine it, since it is making transformations 
into a complete different Fock space)
\begin{eqnarray}
\label{empt}
"emptying"&=& \prod_{\Re \gamma^a}\, \gamma^a \,K =(-)^{\frac{d}{2}+1} \prod_{\Im \gamma^a}\gamma^a \,
\Gamma^{(d)} K\,, 
\end{eqnarray}
and ${\cal C}_{{\cal N}}$
\begin{eqnarray}
\label{CemptPTN}
\mathbb{C}_{{ \cal N}} &=& %\Gamma^{(3+1)}\, \Gamma^{(d)} \,
\prod_{\Re \gamma^a, a=0}^{d} \gamma^a \,\,\,K
\, \prod_{\Im \gamma^m, m=0}^{3} \gamma^m \,\,\,\Gamma^{(3+1)} \,K \,I_{x^6,x^8,\dots,x^{d}}\nonumber\\
&=& %i \,(-)^{\frac{d}{2}}\, 
\prod_{\Re \gamma^s, s=5}^{d} \gamma^s \, \,I_{x^6,x^8,\dots,x^{d}}\,.
% &=&\prod_{\Im \gamma^s, s=5}^{d} \gamma^s \,\,\,K \,I_{x^6,x^8,\dots,x^{d}}\,.
\end{eqnarray}

Let us present also the second quantized notation, following the notation in the ref.~\cite{HNds}.
Let  ${\underline {\bf {\Huge \Psi}}}^{\dagger}_{p}[\Psi_{p}]$ be the creation  operator  creating  
a fermion in the state $\Psi_{p}$  and let ${\mathbf{\Psi}}^{\dagger}_{p}(\vec{x})$
be the second quantized field creating a fermion at position $\vec{x}$. Then 
\begin{eqnarray}
\{ {\underline {\bf {\Huge  \Psi}}}^{\dagger}_{p}[\Psi_{p}] &=& \int \, 
{\mathbf{\Psi}}^{\dagger}_{p}(\vec{x})\, 
\Psi_{p}(\vec{x}) d^{(d-1)} x \,\} 
\,|vac> \nonumber
\end{eqnarray}
so that the antiparticle  state becomes
\begin{eqnarray}
\{ {\underline {\bf \mathbb{C}}}_{{\bf {\cal N}}}\, 
{\underline {\bf {\Huge \Psi}}}^{\dagger}_{p}[\Psi_{p}] &=&  
\int \, {\mathbf{\Psi}}_{p}(\vec{x})\, 
({\cal C}_{{\cal N}} \,\Psi_{p} (\vec{x})) d^{(d-1)} x \} \, |vac> \,.\nonumber
\end{eqnarray}
The antiparticle operator 
${\underline {\bf {\Huge \Psi}}}^{\dagger}_{a}[\Psi_{p}]$,  to the corresponding  particle  
creation operator, can also be written as 
\begin{eqnarray}
\label{makingantip}
{\underline {\bf {\Huge \Psi}}}^{\dagger}_{a}[\Psi_{p}]\, |vac>  &=& 
{\underline {\bf \mathbb{C}}}_{{{\bf \cal N}}}\, 
{\underline {\bf {\Huge \Psi}}}^{\dagger}_{p}[\Psi_{p}]\, |vac>  =  
\int \, {\mathbf{\Psi}}^{\dagger}_{a}(\vec{x})\, 
({\bf \mathbb{C}}_{\cal N}\,\Psi_{p} (\vec{x})) \,d^{(d-1)} x  \, \,|vac> \,,\nonumber\\
{\bf \mathbb{C}}_{\cal H} &=& "emptying"\,\cdot\, {\cal C}_{{\cal H}}  \,.
\end{eqnarray}
%

% TI: To Check Below!

The equations of motion for our toy model (Eqs.~(\ref{weyl},\ref{omegas}), and 
correspondingly the solutions (Eq.~(\ref{mabpsi})) manifest the discrete symmetries 
${\cal C}_{{ \cal N}} \cdot {\cal P}_{{ \cal N}} $, $\mathbb{C}_{{ \cal N}} \cdot $ 
$ {\cal P}_{{ \cal N}} $, ${\cal T}_{{ \cal N}}$ and $\mathbb{C}_{{ \cal N}} \cdot $
${\cal P}_{{ \cal N}} $ $\cdot {\cal T}_{{ \cal N}}$, with the operators presented in 
Eqs.~(\ref{CPTN}, \ref{CemptPTN}). Both, ${\cal C}_{{ \cal N}} \cdot {\cal P}_{{ \cal N}} $
$ \cdot \Psi^{(6)}$  and $\mathbb{C}_{{ \cal N}}$ $\cdot {\cal P}_{{ \cal N}} \Psi^{(6)}$  
(\ref{CemptPTN}) solve the equations of motion, provided that 
$\omega_{56m}(x^0,\vec{x}_{3})$ is a real field. The field $\omega_{56m}(x^0,\vec{x}_{3})$ 
transforms under ${\cal C}_{{ \cal N}} \cdot {\cal P}_{{ \cal N}} $ and 
$\mathbb{C}_{{ \cal N}} $ $\cdot {\cal P}_{{\cal N}}$ to $-\omega_{56}{}^{m}(x^0,-\vec{x}_{3})$,
like the $U(1)$ field must~\cite{ItZu}. We shall comment in the subsect.~\ref{chargemassive} 
that $F$ in Eq.~(\ref{omegas}) transforms into $-F$ for either  ${\cal C}_{{ \cal N}} \cdot$
$ {\cal P}_{{ \cal N}} $ or $\mathbb{C}_{{ \cal N}} \cdot $ $ {\cal P}_{{ \cal N}} $, as well 
as for $\mathbb{C}_{{ \cal N}} \cdot $ ${\cal P}_{{ \cal N}} $ $\cdot {\cal T}_{{ \cal N}}$.

%We shall comment in  subsects.~\ref{chargemassless}, \ref{chargemassive} the solutions of the 
%Weyl equation (\ref{weyl}). In subsect.~\ref{chargelessmassive} the special case with
%the scalar fields $f^{\sigma}_s=f \delta^{\sigma}_{s}$ and $f^{\sigma}_s \omega_{56 \sigma}\,$, 
%$\sigma =((5),(6)), s=(5,6)$, obtaining nonzero vacuum expectation values will be commented. 

%Eq.~(\ref{weyl}) manifests the discrete symmetries  of Eq.~(\ref{CPTN}) and correspondingly also 
%the ${\cal C}_{{ \cal N}} \cdot {\cal P}_{{ \cal N}} $  and 
%$\mathbb{C}_{{ \cal N}} \cdot {\cal P}_{{ \cal N}} $ symmetry: 
%${\cal C}_{{ \cal N}} \cdot {\cal P}_{{ \cal N}} $ ($\gamma^0 \gamma^m p_{0m} 
%+ \gamma^0 \gamma^s p_{0s} $)  $({\cal C}_{{ \cal N}} \cdot {\cal P}_{{ \cal N}})^{-1} $ 
%$= - (\gamma^0 \gamma^m p_{0m} + \gamma^0 \gamma^s p_{0s}) $, while 
%$\mathbb{C}_{{ \cal N}} \cdot {\cal P}_{{ \cal N}} $ ($\gamma^0 \gamma^m p_{0m} + 
%\gamma^0 \gamma^s p_{0s} $) $(\mathbb{C}_{{ \cal N}} \cdot {\cal P}_{{ \cal N}})^{-1} $
%$=\gamma^0 \gamma^m p_{0m} + \gamma^0 \gamma^s p_{0s} $, provided that $\omega^{*}_{56s}$
%$=\omega_{56s}$ for $p_{0s}$ from Eq.~(\ref{weyl}). 

{\it Let us summarize}: The starting action~(\ref{action}) and the corresponding Weyl equation~(\ref{weyl}) 
manifest discrete symmetries from Eqs.~(\ref{CPTN},\ref{CemptPTN}). We comment only the Clifford even 
operators, which keep the transformed states within the starting spinor representation.

\subsection{Massless solutions of the Weyl equation with charges}
\label{chargemassless}
In Eq.~(\ref{Massless}) the massless solution with the spin $\frac{1}{2}$ is presented, solving  
Eq.~(\ref{weyl}) for a toy model with the vielbeins and spin connection fields presented in 
Eqs.~(\ref{omegas}).
For $0 <F\le \frac{1}{2}$ the spinor state is massless, mass protected, and represents, when put on the top 
of the Dirac sea, a free massless charged particle. To simplify the discussions we shall choose 
$F=\frac{1}{2}$.
The state of Eq.~(\ref{mlessF}) solves the Weyl equation
\begin{eqnarray}
\label{weylsimpl}
(-2i S^{03} p^0= p^3)\psi \,,
\end{eqnarray}
where the coordinate system in $d=(3+1)$ was chosen, to simplify the discussions, so that 
$p^{m}=(p^0,0,0,|p^3|)$. 

In Table~\ref{Table I.} (taken from the papers~\cite{TDN,NBled2013,HNds}) all the solutions of the 
Weyl equation are represented. The first two lines represent the two 
positive energy solutions of Eq.~(\ref{weylsimpl}) with the spin $\pm \frac{1}{2}$, both carrying 
the charge $\frac{1}{2}$, both right handed with respect to $d=(3+1)$ and correspondingly mass 
protected.

There are two additional positive energy solutions (the third and the fourth line of the table), 
which manifest indeed the holes in the Dirac sea of the two negative energy 
solutions,  presented in the fifth and the sixth line in Table~\ref{Table I.}. These two positive 
energy solutions - the two holes in the Dirac sea - represent the corresponding antiparticle state 
to the two starting state.

There is the Clifford even discrete symmetry operator ${\cal C}_{{\cal N}}\cdot $ 
$ {\cal P}^{(d-1)}_{{\cal N}}$, Eqs.~(\ref{CPTN},\ref{CemptPTN}) (the operator with the even number 
of $\gamma^a$'s), which transforms
the two positive energy solutions (presented in the first two lines in Table~\ref{Table I.}) into 
the corresponding two negative energy solutions (the first  line transformed into the fifth line 
and the second line into the sixth line), keeping all the states - particles and antiparticles  -
within the starting Weyl representation of Eq.~(\ref{weylrep}). For $d=(5+1)$  the  operator 
${\cal C}_{{\cal N}}\cdot {\cal P}^{(d-1)}_{{\cal N}}$ equals to  $\gamma^0 \gamma^2$ K 
$I_{\vec{x}_3}$ $I_{x^6}$. 

The two antiparticle states of positive energy follow also 
directly from the particle states by the application of the operator 
$\mathbb{C}_{{\cal N}}\cdot {\cal P}^{(d-1)}_{{\cal N}} =$  $\gamma^0 \gamma^5$  
$I_{\vec{x}_3}$ $I_{x^6}$.
 \begin{table}
 \begin{center}
 \begin{tabular}{|c|c|c|c|c|c|r|r|}
 \hline
 $\psi^{(6)}_{i,j}$& positive energy state& 
   $\frac{p^0}{|p^{0}|}$&  $\frac{p^3}{|p^{3}|}$&$(-2iS^{03})$&$\Gamma^{(3+1)}$& $S^{56}$
   &    $\frac{2 p^{3} S^{12}}{|p^0|}$  \\
 \hline
 $\psi^{(6)}_{\frac{1}{2},\frac{1}{2}}$& 
 ${\bf  \stackrel{03}{(+i)}\,\stackrel{12}{(+)}|\stackrel{56}{(+)}\,e^{-i|p^0| x^0 + i|p^3| x^3}}$& 
  $+1$&  $+1$&$+1$&$+1$ & $\frac{1}{2}$ &$1 $ \\
  $\psi^{(6)}_{\frac{1}{2},-\frac{1}{2}}$& 
    $ \stackrel{03}{[-i]}\,\stackrel{12}{[-]}|\stackrel{56}{(+)}\,e^{-i |p^0| x^0 - i|p^3| x^3}$& 
   $+1$&  $-1$&$-1$&$+1$&$\frac{1}{2}$&$1 $ \\
  \hline
 ${\bf \psi^{(6)}_{-\frac{1}{2},\frac{1}{2}}}$& 
  $ \stackrel{03}{[-i]}\,\stackrel{12}{(+)}|\stackrel{56}{[-]}\,e^{-i |p^0| x^0 - i |p^3| x^3}$& 
   $+1$&  $-1$&$-1$&$-1$&$-\frac{1}{2}$&$-1$ \\
   ${\bf \psi^{(6)}_{-\frac{1}{2},-\frac{1}{2}}}$& 
     $ {\bf \stackrel{03}{(+i)}\,\stackrel{12}{[-]}|\stackrel{56}{[-]}\,e^{-i|p^0| x^0 + i |p^3| x^3}}$& 
   $+1$&  $+1$&$+1$&$-1$&$-\frac{1}{2}$&$-1$ \\
 \hline \hline
 $\psi^{(6)}_{i,j}$& negative energy state& 
    $\frac{p^0}{|p^{0}|}$&  $\frac{p^3}{|p^{3}|}$&$(-2iS^{03})$&$\Gamma^{(3+1)}$& $S^{56}$
    &    $\frac{2 p^{3} S^{12}}{|p^0|}$  \\
   \hline
    $\psi^{(6)}_{\frac{1}{2},-\frac{1}{2}}$& 
        $ {\bf \stackrel{03}{[-i]}\,\stackrel{12}{[-]}|\stackrel{56}{(+)}\,e^{i|p^0| x^0 + i|p^3| x^3}}$& 
    $-1$&  $+1$&$-1$&$+1$&$\frac{1}{2}$&$-1$ \\
  $\psi^{(6)}_{\frac{1}{2},\frac{1}{2}}$& 
  $ \stackrel{03}{(+i)}\,\stackrel{12}{(+)}|\stackrel{56}{(+)}\,e^{i|p^0| x^0 - i|p^3| x^3}$& 
   $-1$&  $-1$&$+1$&$+1$&$\frac{1}{2}$&$-1$ \\  
 %$\psi^{(6)}_{\frac{1}{2},\frac{1}{2}}$& 
 %  $ {\bf \stackrel{03}{[-i]}\,\stackrel{12}{[-]}|\stackrel{56}{(+)}\,e^{i|p^0| x^0 + i|p^3| x^3}}$& 
  %  $-1$&  $+1$&$-1$&$+1$&$\frac{1}{2}$&$-1$ \\  
  %$ \psi^{neg}_{2}$& 
  % $\stackrel{03}{(+i)}\,\stackrel{12}{[-]}|\stackrel{56}{[-]}\,e^{i|p^0| x^0 - i|p^3| x^3}$& 
  %  $-1$&  $-1$&$+1$&$-1$&$-\frac{1}{2}$&$1 $ \\
  
  %${\bf \psi^{neg}_{4}}$& 
  % $ {\bf \stackrel{03}{[-i]}\,\stackrel{12}{(+)}|\stackrel{56}{[-]}\,e^{i|p^0| x^0 + i|p^3| x^3}}$& 
  %  $-1$&  $+1$&$-1$&$-1$&$-\frac{1}{2}$&$1 $ \\
 \hline
 \end{tabular}
 \end{center}
 \caption{\label{Table I.} 
 Two positive energy states of the charge $\frac{1}{2}$ (index i), the right handed (with respect 
 to $d=(3+1)$) and  with the spin  (determined by the index $j$) $\frac{1}{2}$ , the first line, 
 and $-\frac{1}{2}$, the second line, representing  particles when put on the  top of the Dirac sea. 
 The operator  ${\cal C}_{{\cal N}}\cdot {\cal P}^{(d-1)}_{{\cal N}} =$ $\gamma^0 \gamma^2$ K 
 $I_{\vec{x}_3}$ $I_{x^6}$ 
 transforms these two states into the negative energy state in the Dirac sea, presented in the last two 
 lines of the table; the first line into the fifth one and the second into the sixth one. 
 The remaining two positive energy states with the charge $-\frac{1}{2}$ represent holes in the 
 Dirac sea, the third line corresponds to the fifth and the fourth line to the sixth one.  
 These are the two antiparticle states, put on the top of the Dirac sea, which follow also directly 
 from the starting particle state by the application of 
 $\mathbb{C}_{{ \cal N}}\cdot {\cal P}^{(d-1)}_{{\cal N}} =$ $\gamma^0 \gamma^5$ 
 $I_{\vec{x}_3}$ $I_{x^6}$. The coordinate system in $d=(3+1)$ is chosen so that $p^m=(p^0,0,0,p^3)$.
 $\Gamma^{(5+1)}=-1$ and $\Gamma^{((d-1)+1)}$  define the handedness in $d=(5+1)$-dimensional space-time,
 $S^{56}$ defines the charge in $d=(3+1)$, $\frac{2 p^{3} S^{12}}{|p^0|}$ defines the helicity. 
 Nilpotents $ \stackrel{ab}{(k)}$ and  projectors $ \stackrel{ab}{[k]}$  operate on the vacuum 
 state  $|vac>_{fam}$ not written in the table. Table is partly taken from~\cite{NBled2013,TDN,HNds}.    
}
 \end{table}

The states, presented in Table~\ref{Table I.}, are the solutions of the Weyl equations
\begin{eqnarray}
\label{weylhh}
(\Gamma^{(3+1)} \frac{p^0}{|p^0|}= \frac{2 \vec{p}\cdot \vec{S}}{|p^0|})\,\psi\,.
\end{eqnarray}
for the choice 
%$\vec{p}=(p^1,p^2,p^3)$, in our case is 
$(0,0,p^3)$. %, presented in all text books. 
Here $\vec{S}= (S^{23}, S^{31}, S^{12})$, $S^{ab}=\frac{i}{2} (\gamma^a \gamma^b- \gamma^b\gamma^a)$,
and $\Gamma^{((d-1)+1)}$ (in usual notation is for $d=(3+1)$ named $\gamma^5$) determines handedness
for fermions in any $d$. For $d=(5+1)$, $\Gamma^{(5+1)}= \prod_{a}\,\gamma^a$ in ascending order,  equal 
also to $\Gamma^{(3+1)} \cdot (-2 S^{56})$. 

For the choice $p^m=(p^1,p^2,p^3)$ and the spin $\pm \frac{1}{2}$ are the solutions presented in 
Eq.~(\ref{weylgen}), the two particle states of positive energy are in the first two lines, the 
corresponding particle states of the negative energy are in the last two lines, while  the 
corresponding antiparticle states, representing the hole in the negative energy states of the Dirac 
sea are written in the third and the fourth line. As in the simplified case also in this general case 
the negative energy states and the antiparticle states follow from the positive energy states by
the application of ${\cal C}_{{ \cal N}} \cdot {\cal P}_{{ \cal N}} $ and $\mathbb{C}_{{ \cal N}} \cdot $ 
$ {\cal P}_{{ \cal N}} $, respectively. 
\begin{eqnarray}
\label{weylgen}
&& {\rm particle} \; {\rm states} \nonumber\\
p^0&=&|p^0|\,,\nonumber\\
\psi^{(6)}_{\frac{1}{2},\frac{1}{2}}\,(\vec{p})  &=& \left( \stackrel{03}{(+i)}\,\stackrel{12}{(+)}|
\stackrel{56}{(+)}
+ \frac{p^1 +i p^2}{ |p^0| + |p^3|} \stackrel{03}{[-i]}\,\stackrel{12}{[-]}|\stackrel{56}{(+)}\right)\,
e^{-i(|p^0| x^0 - \vec{p}\cdot\vec{x})}\,,\nonumber\\
\psi^{(6)}_{\,\frac{1}{2},-\frac{1}{2}}\,(\vec{p})&=& \left(\stackrel{03}{[-i]}\,\stackrel{12}{[-]}|
\stackrel{56}{(+)}
- \frac{p^1 -i p^2}{ |p^0| + |p^3|} \stackrel{03}{(+i)}\,\stackrel{12}{(+)}|\stackrel{56}{(+)}\right)\,
e^{-i(|p^0| x^0 + \vec{p}\cdot\vec{x})}\,,\nonumber\\
&& {\rm antiparticle} \; {\rm states} \nonumber\\
\psi^{(6)}_{-\frac{1}{2},\,\frac{1}{2}}\,(\vec{p})  &=& \left( \stackrel{03}{[-i]}\,\stackrel{12}{(+)}|
\stackrel{56}{[-]}
+ \frac{p^1 +i p^2}{ |p^0| + |p^3|} \stackrel{03}{(+i)}\,\stackrel{12}{[-]}|\stackrel{56}{[-]}\right)\,
e^{-i(|p^0| x^0 + \vec{p}\cdot\vec{x})}\,,\nonumber\\
\psi^{(6)}_{-\frac{1}{2},-\frac{1}{2}}\,(\vec{p})&=& \left(\stackrel{03}{(+i)}\,\stackrel{12}{[-]}|
\stackrel{56}{[-]}
- \frac{p^1 -i p^2}{ |p^0| + |p^3|} \stackrel{03}{[-i]}\,\stackrel{12}{(+)}|\stackrel{56}{[-]}\right)\,
e^{-i(|p^0| x^0 - \vec{p}\cdot\vec{x})}\,,\nonumber\\
&& {\rm states } \; {\rm in} \; {\rm the}\; {\rm Dirac}\;{\rm sea} \nonumber\\
p^0&=&-|p^0|\,,\nonumber\\
\psi^{(6)}_{\,\frac{1}{2},-\frac{1}{2}}\,(\vec{p})&=& \left(\stackrel{03}{[-i]}\,\stackrel{12}{[-]}|
\stackrel{56}{(+)}
- \frac{p^1 - i p^2}{ |p^0| + p^3} \stackrel{03}{(+i)}\,\stackrel{12}{(+)}|\stackrel{56}{(+)}\right)\,
e^{i(|p^0| x^0 + \vec{p}\cdot\vec{x})}\,,\nonumber\\
\psi^{(6)}_{\,\frac{1}{2},\,\frac{1}{2}}(\vec{p})&=&  \left(\stackrel{03}{(+i)}\,\stackrel{12}{(+)}|
\stackrel{56}{(+)} + \frac{p^1 +i p^2}{ |p^0| + p^3} \stackrel{03}{[-i]}\,\stackrel{12}{[-]}|
\stackrel{56}{(+)}\right)\,e^{i(|p^0| x^0 - \vec{p}\cdot\vec{x})}\,,\nonumber\\
\end{eqnarray}
We do not pay attention on the normalization.

{\it Let us summarize}: We presented massless solutions of the Weyl equation~(Eqs.(\ref{weyl},\ref{omegas})) 
for the positive and the negative energies. The positive energy states, put on the top of the Dirac sea, 
represent particles. The negative energy states can be found also from the positive energy states by 
the application of ${\cal C}_{{ \cal N}}\cdot {\cal P}^{(d-1)}_{{\cal N}} =$ $\gamma^0 \gamma^2$ K 
$I_{\vec{x}_3}$ $I_{x^6}$. Antiparticle states follow from the particle ones by either
emptying the negative energy state in the Dirac sea or directly from the particle states by the 
application of the operator  $\mathbb{C}_{{ \cal N}}\cdot {\cal P}^{(d-1)}_{{\cal N}} =$ 
$\gamma^0 \gamma^5$  $I_{\vec{x}_3}$ $I_{x^6}$ and putting these states on the top of the Dirac sea.
There are, for each choice of the four momentum ($|p^0|, p^1,p^2,p^3$) two positive and two negative 
energy states, while the two antiparticle states represent the hole in the Dirac sea.
The antiparticle states, having opposite charges than the corresponding particle states, have obviously 
also different handedness $\Gamma^{3+1}$ - but still left handedness
with respect to $\gamma^{5+1}$ - than the particle states. 

\subsection{Massive solutions of the Weyl equation with  charges} 
\label{chargemassive}

Since the discrete symmetries of Eqs.(\ref{CPTN}, \ref{CemptPTN}) are the symmetries of the 
equations of motion (\ref{weyl}), also the massive states, presented in Eq.(\ref{mabpsi}), 
manifest these symmetries. As discussed in the ref.~\cite{HNds}, and can easily be checked, the 
state with the charge $(n+\frac{1}{2})$ and spin $\frac{1}{2}$, $\,\psi^{(6)}_{n+\frac{1}{2}, \frac{1}{2}}$,  
which has  for $F=\frac{1}{2}$ the mass $(m \rho_0)^2=l(l+1)$, 
$l=1,2,3,\dots$; $0 \le n \le l$,  transforms under 
$\mathbb{C}_{{\cal N}}\cdot {\cal P}^{(d-1)}_{{\cal N}}$ into its anti-particle state 
$\psi^{(6)}_{-n-\frac{1}{2},\frac{1}{2}}$ of the same spin and mass and opposite charge
\begin{eqnarray}
&&\mathbb{C}_{{\cal N}}\cdot {\cal P}^{(d-1)}_{{\cal N}} \psi^{(6) }_{n+\frac{1}{2}, \frac{1}{2}}= 
({\cal B}_{n+1}\,\stackrel{03}{(+i)} \stackrel{12}{(+)} \stackrel{56}{(+)}  
+ {\cal A}_{n}\, e^{i \phi}\, \stackrel{03}{[-i]}\stackrel{12}{(+)}\,\stackrel{56}{[-]})\,
\cdot e^{-i(n+1) \phi} e^{-i(p^0 x^0+ p^3 x^3)}\,. 
\label{mabpsianti}
\end{eqnarray}
This massive antiparticle state (if put on the top of the Dirac sea,  representing the 
hole in the Dirac sea in the negative energy state ${\cal C}_{{\cal N}}\cdot $ 
${\cal P}^{(d-1)}_{{\cal N}} $ $\psi^{(6) }_{n+1/2,\frac{1}{2}}$) solves the Weyl 
equation~(\ref{equationm56})  with $F\rightarrow -F$, 
${\cal B}_{n+1}= {\cal A}_{-n-1}$ and ${\cal A}_{n}= {\cal B}_{-n}$. 

The massive particles carry the charges $(n+\frac{1}{2})$ and the masses $(m \rho_0)^2=l(l+1)$, 
$l=1,2,3,\dots$; $0 \le n \le l$, which depend on the radius of $S^2$. The corresponding 
antiparticles carry the same mass (determined by $l$) and the opposite charge $(-n-\frac{1}{2})$.

{\it Let us summarize}: There are two positive energy states for each mass $(m \rho_0)^2=l(l+1)$ 
and ($|p^0|, p^1,p^2,p^3$) and each charge $n+\frac{1}{2}$ with the spin $\pm \frac{1}{2}$, 
which, if put on the top of the Dirac sea, represent the particles. There are two corresponding 
negative energy states and there are the corresponding two antiparticle states of different 
handedness  $\Gamma^{3+1}$ (in the usual notation $\gamma^5$), which, put on the top of the 
Dirac sea, represent the holes in the Dirac sea.

\subsection{Massive chargeless solutions of the Weyl equation} 
\label{chargelessmassive}

Let us now assume that the scalar fields, the gauge fields of $S^{56}$, that is  $f^{\sigma}_{s}$ 
$ \omega_{56 \sigma}$, with $s=(5,6)$ and $\sigma=((5),(6))$, gain  non zero vacuum expectation 
values. These two scalar fields are the analogy to the complex higgs scalar of the {\em standard 
model}: Higgs in the {\it standard model} carries the weak and the hyper charge, while our scalar 
fields carry  only the "hyper" charge $S^{56}$.  The charge, which is the spin in $d=(5,6)$,  
is after the scalar fields gain nonzero vacuum expectation values no longer the conserved quantity. 

In this case we replace in the Weyl equation~(\ref{weyl}) the quantities  $p_{0\pm}$ with 
their  %~\cite{DHN}  
vacuum expectation values $<p_{0\pm}>$, so that the equations of motion read 
 \begin{eqnarray}
 \label{Weylmvac}
 &&<p_{0\pm}> = m_{\pm} \,,\nonumber\\
 &&(\gamma^m p_{m}  + \stackrel{56}{(+)} \,m_{+}  + \stackrel{56}{(-)}\, m_{-} )\, \psi^{(6)}=0
 \,.\nonumber\\
 \end{eqnarray}
To simplify, the coordinate system in $d=(3+1)$  with  $\vec{p}=0$ is chosen. 
Then  Eq.~(\ref{Weylmvac}) reads 
 \begin{eqnarray}
 \label{Weylmvacsimple}
  &&\{p_{0} + \gamma^0 (\stackrel{56}{(+)} \,m_{+}  + \stackrel{56}{(-)}\, m_{-} )\} \psi^{(6)}=0
  \,,\nonumber\\
 %p_{0\pm}= p_{0}^5 \mp i\,p_{0}^6\,,\quad p_{0s}=  f^{\sigma}_{s}(p_{\sigma} -\frac{1}{2} S^{ab}
 %\omega_{ab \sigma})= p_{s}-  S^{56}i \omega_{56 s}\,.
 %- \frac{1}{2}  \tilde{S}^{ab}\tilde{\omega}_{ab \sigma})
 \end{eqnarray}
 %
 %There is no conserved charge any longer for these massive states with the nonzero mass $m$. 
The two positive and the two negative energy solutions with the spin in $d=(3+1)$ $\pm \frac{1}{2}$, 
%$\psi^{(6)}_{\pm\frac{1}{2},m}$, 
the nonzero mass $m$ manifesting in $d=(3+1)$ (with no conserved charge) obeying 
Eq.~(\ref{Weylmvacsimple}), are as follows 
 \begin{eqnarray}
 \label{Weylmsolsimple}
\psi^{(6)}_{\;\frac{1}{2},m}&=& (\stackrel{03}{(+i)} \stackrel{12}{(+)} \stackrel{56}{(+)} + 
\frac{m\;}{m_{+}} \,\stackrel{03}{[-i]} \stackrel{12}{(+)} \stackrel{56}{[-]})\, e^{-imx^0} \,, \nonumber\\
\psi^{(6)}_{-\frac{1}{2},m}&=& (\stackrel{03}{[-i]} \stackrel{12}{[-]} \stackrel{56}{(+)} + 
\frac{m\;}{m_{+}} \stackrel{03}{(+i)} \stackrel{12}{[-]} \stackrel{56}{[-]})\, e^{-imx^0} \,, \nonumber\\ 
{\cal C}_{{\cal N}}\cdot {\cal P}^{(d-1)}_{{\cal N}}\,\,\psi^{(6)}_{\;\frac{1}{2},m}&=& 
-i \,(\stackrel{03}{[-i]} \stackrel{12}{[-]} \stackrel{56}{(+)} -
 \frac{m\;}{m_{+}}\, \stackrel{03}{(+i)} \stackrel{12}{[-]} \stackrel{56}{[-]})\, e^{imx^0} \,, \nonumber\\
 {\cal C}_{{\cal N}}\cdot {\cal P}^{(d-1)}_{{\cal N}}\,\,\psi^{(6)}_{-\frac{1}{2},m}&=& 
\,i\,(\stackrel{03}{(+i)} \stackrel{12}{(+)} \stackrel{56}{(+)}  - \frac{m\;}{m_{+}}
 \stackrel{03}{[-i]} \stackrel{12}{(+)} \stackrel{56}{[-]})\, e^{imx^0} \,,\nonumber\\
 m^2&=& m_{+}\, m_{-}\,,\quad m_{+} = -m_{-}\,, \quad (p_{0})^2 = m^2\,.
\end{eqnarray}
There is no massless state any longer.

Let us notice that  $\,({\cal C}_{{\cal N}}\cdot {\cal P}^{(d-1)}_{{\cal N}})$ $\,p_{0 \pm}\,$
$({\cal C}_{{\cal N}}\cdot {\cal P}^{(d-1)}_{{\cal N}})^{-1}$ $=-p_{0 \pm}$ (provided that 
$\omega^{*}_{56 \pm}= \omega_{56 \pm}$). Accepting the expectation values  $<p_{0\pm}>$ as the "mass" 
term $m_{\pm}$  (indeed $m^2= m_{+} m_{-}$) it then follows that $\,({\cal C}_{{\cal N}}\cdot $ $
{\cal P}^{(d-1)}_{{\cal N}})$ $\,m_{\pm}\,$ $({\cal C}_{{\cal N}}\cdot {\cal P}^{(d-1)}_{{\cal N}})^{-1}$ 
$=-m_{\pm}= m_{\mp}$, requiring that both, $m_{\pm}$, are imaginary, in accordance with 
Eq.~(\ref{Weylmsolsimple}). 
Since also ${\cal C}_{{\cal N}}\cdot {\cal P}^{(d-1)}_{{\cal N}}$ $\gamma^0 \,$ $\stackrel{56}{(\pm)}$
$({\cal C}_{{\cal N}}\cdot {\cal P}^{(d-1)}_{{\cal N}})^{-1}=$ $\gamma^0 \,$ $\stackrel{56}{(\pm)}$, 
${\cal C}_{{\cal N}}\cdot {\cal P}^{(d-1)}_{{\cal N}}$ stays as the symmetry  of the effective 
equations of  motion.

Let us check also the discrete symmetry operator  
$\mathbb{C}_{{\cal N}}\cdot {\cal P}^{(d-1)}_{{\cal N}}$. 
One finds that $\mathbb{C}_{{\cal N}}\cdot {\cal P}^{(d-1)}_{{\cal N}}$ $\gamma^0 \,$ 
$\stackrel{56}{(\pm)}$ $(\mathbb{C}_{{\cal N}}\cdot {\cal P}^{(d-1)}_{{\cal N}})^{-1}=$ 
$\gamma^0 \,\stackrel{56}{(\mp)}$ and that 
$\mathbb{C}_{{\cal N}}\cdot {\cal P}^{(d-1)}_{{\cal N}}$ $p_{0\pm}$
$(\mathbb{C}_{{\cal N}}\cdot {\cal P}^{(d-1)}_{{\cal N}})^{-1}=$ $p_{0\mp}$, while 
$\mathbb{C}_{{\cal N}}\cdot {\cal P}^{(d-1)}_{{\cal N}}$ $m_{\pm}$
$(\mathbb{C}_{{\cal N}}\cdot {\cal P}^{(d-1)}_{{\cal N}})^{-1}=$ $m_{\pm}$. This means that
the effective Weyl equation (Eq.\ref{Weylmvacsimple}) is not invariant with respect to 
$\mathbb{C}_{{\cal N}}\cdot {\cal P}^{(d-1)}_{{\cal N}}$.  For this effective Weyl equation the 
operator $"emptying"$ must be therefore changed. Let us multiply it with $( -i\,\Gamma^{(6)}$ 
$\Gamma^{(3+1)})$. The operator $\Gamma^{(6)}$ is a number, which is ($-1$) in the case of our 
starting representation, Eq.~(\ref{weylrep}), while $\Gamma^{(3+1)}$  distinguishes among 
these four states, having the alternating values $\pm 1$, respectively. One finds that 
$ -i\,\Gamma^{(6)}$ $\Gamma^{(3+1)}=$ $\gamma^5\,\gamma^6$ ($= -2i S^{56}$).  
 Let us now check how does the equation of motion $\{\gamma^0 \gamma^m p_{m} + \gamma^0 
 (\stackrel{56}{(+)} m_{+} + \stackrel{56}{(-)} m_{-})\}\,\psi^{(6)}=0$ 
transform with respect to this new proposed symmetry 
 \begin{eqnarray}
 \label{CPNeffnew}
  &&(-i)\,\Gamma^{(6)}\Gamma^{(3+1)}\,\mathbb{C}_{{\cal N}}\cdot {\cal P}^{(d-1)}_{{\cal N}}\;
  (\gamma^0\,\gamma^m p_{m}  + \gamma^0\,(\stackrel{56}{(+)} \,m_{+}  + \stackrel{56}{(-)}\, m_{-} ))\; 
 ((-i)\,\Gamma^{(6)} \Gamma^{(3+1)}\,\mathbb{C}_{{\cal N}}\cdot {\cal P}^{(d-1)}_{{\cal N}})^{-1} 
 \nonumber\\
 && =(\gamma^0\,\gamma^m p_{m}  + \gamma^0\,((-) \stackrel{56}{(-)} \,m_{+}  + 
 (-)\stackrel{56}{(+)}\, m_{-}))= 
 (\gamma^0\,\gamma^m p_{m}  + \gamma^0\,(\stackrel{56}{(-)} \,m_{-}  + 
 (\stackrel{56}{(+)}\, m_{+}))\,, \nonumber\\
 &&{\rm since} \quad m_{+}= - m_{-}\,.
 %p_{0\pm}= p_{0}^5 \mp i\,p_{0}^6\,,\quad p_{0s}=  f^{\sigma}_{s}(p_{\sigma} -\frac{1}{2} S^{ab}
 %\omega_{ab \sigma})= p_{s}-  S^{56}i \omega_{56 s}\,.
 %- \frac{1}{2}  \tilde{S}^{ab}\tilde{\omega}_{ab \sigma})
 \end{eqnarray}
We see that the effective Weyl equation (Eq.~\ref{Weylmvacsimple}) is invariant under this new  
discrete symmetry operator $ -i\,\Gamma^{(6)}$ $\Gamma^{(3+1)}$ 
$\mathbb{C}_{{\cal N}}\cdot {\cal P}^{(d-1)}_{{\cal N}}$ ($= \gamma^0 \gamma^6 I_{\vec{x}_{3}} 
I_{x^6}$), which again first transforms the positive energy state (the particle if the state is 
put on the top of the Dirac sea) into the corresponding  negative energy state and then empties 
it so that we have now the antiparticle state put on the top of the Dirac sea.

Let us calculate the two positive energy antiparticle states (when put on the top of the Dirac sea),
missing in Eq.~(\ref{Weylmsolsimple})
\begin{eqnarray}
 \label{Weylmsolsimpleanti}
 -i\,\Gamma^{(6)} \Gamma^{(3+1)} \mathbb{C}_{{\cal N}}\cdot {\cal P}^{(d-1)}_{{\cal N}} \;
\psi^{(6)}_{\;\frac{1}{2},m}&=& ( (-i)\stackrel{03}{[-i]} \stackrel{12}{(+)} \stackrel{56}{[-]} + 
\frac{m\;}{m_{+}}\,i \,\stackrel{03}{(+i)} \stackrel{12}{(+)} \stackrel{56}{(+)})\, e^{-imx^0} \,, \nonumber\\
-i\,\Gamma^{(6)} \Gamma^{(3+1)} \mathbb{C}_{{\cal N}}\cdot {\cal P}^{(d-1)}_{{\cal N}} 
\psi^{(6)}_{-\frac{1}{2},m}&=& ((-i)\stackrel{03}{(+i)} \stackrel{12}{[-]} \stackrel{56}{[-]} + 
i \,\frac{m\;}{m_{+}} \stackrel{03}{[-i]} \stackrel{12}{[-]} \stackrel{56}{(+)})\, e^{-imx^0} \,,
\nonumber\\
 m^2&=& m_{+}\, m_{-}\,,\quad m_{+} = -m_{-}\,, \quad (p_{0})^2 = m^2\,,
\end{eqnarray}
which means that $-i\,\Gamma^{(6)} \Gamma^{(3+1)}$ $ \mathbb{C}_{{\cal N}}\cdot $ 
${\cal P}^{(d-1)}_{{\cal N}}\, \psi^{(6)}_{\pm \,\frac{1}{2},m}$ $=i \frac{m\;}{m_{+}}\,$ $
\psi^{(6)}_{\pm \,\frac{1}{2},m}$. The two positive solutions of the effective Weyl equations 
(Eq.~\ref{Weylmvacsimple})  representing particle carrying no charge  are indistinguishable from 
the two positive energy solutions for the corresponding two antiparticles, which are indeed the 
two holes in the Dirac sea.

One sees that the other two superposition, $\psi^{'(6)}_{\;\frac{1}{2},m} = $ $(\stackrel{03}{(+i)} 
\stackrel{12}{(+)} \stackrel{56}{(+)} + \frac{m\;}{m_{-}} $ $\stackrel{03}{[-i]} \stackrel{12}{(+)} $
$\stackrel{56}{[-]})\,$ $ e^{-imx^0} \,$ and $\psi^{'(6)}_{-\frac{1}{2},m} = $ $(\stackrel{03}{[-i]} $
$\stackrel{12}{[-]} \stackrel{56}{(+)} \; + \frac{m\;}{m_{-}} $ $ \stackrel{03}{(+i)} $ 
$\stackrel{12}{[-]} \stackrel{56}{[-]})\,$ $ e^{-imx^0} $ are not   solutions of the Weyl 
equation and so are not also the negative energy states, obtained by the application of 
${\cal C}_{{\cal N}}\cdot $ ${\cal P}^{(d-1)}_{{\cal N}}$ on these two states. 

In this discussion only one family is assumed.
To obtain true masses of spinors one must take into account the families and also the 
loop corrections in all orders, to which also the dynamical  scalar and vector gauge fields 
contribute. 

{\it Let us summarize}: 
In our effective equations of motion  (Eq.~(\ref{Weylmvac})) the vacuum expectation values of the 
scalar fields are assumed. Correspondingly no massless solutions exist any longer. We are left instead 
with  the
two positive  energy states (representing particles if put on the top of the Dirac sea) and the 
two corresponding negative energy states, the holes in which represent (put on the top of the 
Dirac sea) the antiparticle states (Eq.~(\ref{Weylmsolsimpleanti})). The antiparticle states are 
indistinguishable from the particle states and are indeed the Majorana particles.  For the effective  
equations of motion the discrete symmetry "emptying" has to be changed, so that the operator which 
transform the particle state into the antiparticle state is $-i\,\Gamma^{(6)} \Gamma^{(3+1)}$ 
$ \mathbb{C}_{{\cal N}}\cdot $ 
${\cal P}^{(d-1)}_{{\cal N}}$.

All these states solve the Weyl equation (\ref{Weylmvacsimple}).

\subsection{Dirac representation of spinor states} 
\label{Dirac} 

Let us connect our positive and  negative  energy solutions for the particular choice of the 
coordinate system ($p^m=(m,0,0,0)$) and then for the general choice of the coordinate 
system $p^m=(p^0,p^1,p^2,p^3)$, solving the Weyl equation (Eq.~\ref{Weylmvac})
\begin{eqnarray}
\label{mord}
p^0= |p^0|\,, \nonumber\\
\psi^{(6)}_{\;\frac{1}{2},m}(\vec{p})&=& {\cal N}(\vec{p},m)\,
\{  \stackrel{03}{(+i)} \stackrel{12}{(+)} \stackrel{56}{(+)} + \frac{m\;}{m_{+}}\, 
\frac{|p^0|-|p^3| +m}{|p^0|+ |p^3| +m}\,
       \stackrel{03}{[-i]} \stackrel{12}{(+)} \stackrel{56}{[-]} \nonumber\\
&+& \frac{p^1 +i p^2}{|p^0|+ |p^3| +m}\,
      (\stackrel{03}{[-i]} \stackrel{12}{[-]} \stackrel{56}{(+)}
    -\frac{m\;}{m_{+}} \, \stackrel{03}{(+i)} \stackrel{12}{[-]} \stackrel{56}{[-]})
\}\,\cdot e^{-i(|p^0|x^0-\vec{p}\cdot\vec{x})}\,
\nonumber\\
\psi^{(6)}_{-\frac{1}{2},m}(\vec{p})&=& {\cal N}(\vec{p},m)\,
\{  \frac{p^1 -i p^2}{|p^0|+ |p^3| +m}\,
(\stackrel{03}{(+i)} \stackrel{12}{(+)} \stackrel{56}{(+)} - \frac{m\;}{m_{+}} \,
       \stackrel{03}{[-i]} \stackrel{12}{(+)} \stackrel{56}{[-]}) \nonumber\\
&+& \frac{|p^0|-|p^3|+m}{|p^0|+ |p^3| +m}\cdot
     \stackrel{03}{[-i]} \stackrel{12}{[-]} \stackrel{56}{(+)}
   + \frac{m\;}{m_{+}} \, \stackrel{03}{(+i)} \stackrel{12}{[-]} \stackrel{56}{[-]}
\}\,\cdot e^{-i(|p^0|x^0-\vec{p}\cdot\vec{x})}\,,
\nonumber\\
p^0= -|p^0|\,, \nonumber\\
\psi^{(6)}_{\;\frac{1}{2},m}(\vec{p})&=& {\cal N}(\vec{p},m)\,
\{ \frac{p^1 + i p^2}{|p^0|+ |p^3| +m}\,
 (\stackrel{03}{[-i]} \stackrel{12}{[-]} \stackrel{56}{(+)} + \frac{m\;}{m_{+}} 
    \stackrel{03}{(+i)} \stackrel{12}{[-]} \stackrel{56}{[-]})\nonumber\\
&+& (-)(\frac{|p^0|-|p^3|+m}{|p^0|+|p^3| +m}\,
   \stackrel{03}{(+i)} \stackrel{12}{(+)} \stackrel{56}{(+)}
- \frac{m\;}{m_{+}} \,  \stackrel{03}{[-i]} \stackrel{12}{(+)} \stackrel{56}{[-]})
\}\,\cdot e^{i(|p^0|x^0+\vec{p}\cdot\vec{x})}\,,\nonumber\\
\psi^{(6)}_{-\frac{1}{2},m}(\vec{p})&=& {\cal N}(\vec{p},m)\,
\{  \stackrel{03}{[-i]} \stackrel{12}{[-]} \stackrel{56}{(+)} -\frac{m\;}{m_{+}}\,
\frac{|p^0|-|p^3| +m}{|p^0|+ |p^3| +m}\,
    \stackrel{03}{(+i)} \stackrel{12}{[-]} \stackrel{56}{[-]}\nonumber\\
&+& (-)\,\frac{p^1 -i p^2}{|p^0|+ |p^3| +m}\,
   (\stackrel{03}{(+i)} \stackrel{12}{(+)} \stackrel{56}{(+)}
+ \frac{m\;}{m_{+}}\, \stackrel{03}{[-i]} \stackrel{12}{(+)} \stackrel{56}{[-]})
\}\,\cdot e^{i(|p^0|x^0+\vec{p}\cdot\vec{x})}\,,
\end{eqnarray}
with the usual notation in the textbooks~\cite{ItZu}. 

The two positive and the two negative solutions of Eq.~(\ref{Weylmsolsimple}), 
($\psi^{(6)}_{\frac{1}{2},m}$, $\psi^{(6)}_{-\frac{1}{2},m}$ with $ p^0=|p^0|$ and 
$i\, {\cal C}_{{\cal N}}\cdot {\cal P}^{(d-1)}_{{\cal N}}\, \psi^{(6)}_{\frac{1}{2},m}$, 
$-i\, {\cal C}_{{\cal N}}\cdot {\cal P}^{(d-1)}_{{\cal N}}\psi^{(6)}_{-\frac{1}{2},m}$ 
with $ p^0=-|p^0|$), respectively, if we do not pay attention on normalization, represent the 
four states, usually written as
\begin{equation}\label{ussualpossol}
\begin{pmatrix}\varphi \\ \frac{\vec{\sigma}\cdot \vec{p}}{|p^0|+m}\varphi \end{pmatrix}\,
e^{-i(p^0 x^0-\vec{p}\cdot\vec{x})}\,,
\qquad \varphi=\begin{pmatrix} 1 \\ 0 \end{pmatrix}\, \; {\rm or} 
\qquad \varphi=\begin{pmatrix} 0 \\ 1 \end{pmatrix}\,,\nonumber\\
\end{equation}
\begin{equation}\label{ussualnegsol}
\begin{pmatrix} \frac{\vec{\sigma}\cdot \vec{p}}{|p^0|+m} \chi \\
  \chi\end{pmatrix}\, e^{i(p^0 x^0-\vec{p}\cdot\vec{x})}\,,  
\qquad  \chi=\begin{pmatrix} 1 \\ 0 \end{pmatrix}\,\; {\rm or} 
\qquad  \chi=\begin{pmatrix} 0 \\ 1 \end{pmatrix}\,,
\end{equation}
with $\frac{\vec{\sigma}}{2}= (S^{23}, S^{31}, S^{12})$. In Eq.~(\ref{Weylmsolsimple})  $p^0=m$ and 
$ \vec{p}=0$, while the general case (Eq.~(\ref{mord})) corresponds to $p^{m}= (p^0, p^1,p^2,p^3)$.

The two positive energy antiparticle states, put on the top of the Dirac sea, obtained  from 
the two positive energy states in Eq.~(\ref{mord}) by the application of 
$-i\,\Gamma^{(6)} \Gamma^{(3+1)} \mathbb{C}_{{\cal N}}\cdot {\cal P}^{(d-1)}_{{\cal N}} \;$ 
(or by emptying the corresponding negative energy states),
are the same as the starting two particle states, if $\vec{p}$ is replaced by $-\vec{p}$, in 
the exponent ($e^{-i|p|^0 \pm i\vec{p}\cdot \vec{x}}$ into $e^{-i|p|^0 \mp i\vec{p}\cdot \vec{x}}$),
and in all the coefficients of the wave functions ($(p^1,p^2,|p^3|)$ go to $(-p^1,-p^2,-|p^3|)$)
%similarly  as in the case when $p^m=(m,0,0,0)$
%
\begin{eqnarray}
\label{Weylmsolgenanti}
&&i\,\Gamma^{(6)} \Gamma^{(3+1)} \mathbb{C}_{{\cal N}}\cdot {\cal P}^{(d-1)}_{{\cal N}} \;
\psi^{(6)}_{\;\pm\,\frac{1}{2},m, \vec{p}} \, \rightarrow  \,\psi^{(6)}_{\;\pm \frac{1}{2},m, -\vec{p}}
\,, \nonumber\\
%&&i\,\Gamma^{(6)} \Gamma^{(3+1)} \mathbb{C}_{{\cal N}}\cdot {\cal P}^{(d-1)}_{{\cal N}} 
%\psi^{(6)}_{-\frac{1}{2},m}= - i \frac{m\;}{m_{+}} \,\psi^{(6)}_{-\frac{1}{2},m}\,,
%\nonumber\\
&& m^2= m_{+}\, m_{-}\,,\quad m_{+} = -m_{-}\,, \quad (p_{0})^2 =\vec{p}^{2}+ m^2\,.
\end{eqnarray}

{\it Let us summarize}: 
The two particle and two antiparticle states with spin either $\frac{1}{2}$ or 
$-\frac{1}{2}$ are for $\vec{p}=0$ in $d=(3+1)$ undistinguishable, due to the fact that there are 
no conserved charges. For $\vec{p}\ne 0$, the particle and antiparticle state distinguish in the 
sign of $\vec{p}$. This means that in Eq.~(\ref{ussualnegsol}) $\vec{p}$ for particles must  be 
replaced by $-\vec{p}$. This means again that the particle states are indistinguishable from the 
antiparticle ones. 

%There are correspondingly two positive and two negative energy states, which are the solutions 
%of the equations of motion.

%
\subsection{Majorana spinors} 
\label{Majorana} 

In the case that there is no conserved charge the Majorana particle, the state of which (put on
the top of the Dirac sea) is the sum of the particle and the corresponding antiparticle 
state~(\ref{Weylmsolgenanti}), reads for $\vec{p}=0$ 
\begin{eqnarray}
\label{Majoranastates}
\psi^{(6)}_{\pm \frac{1}{2},\tiny{majorana}} &=&\frac{1}{\sqrt{2}} \,(\psi^{(6)}_{\pm \frac{1}{2},m}\,
+(\pm)\, (i)\,\Gamma^{(6)} \Gamma^{(3+1)} \mathbb{C}_{{\cal N}}\cdot {\cal P}^{(d-1)}_{{\cal N}} \;
\psi^{(6)}_{\pm \frac{1}{2},m}(\vec{p}=0)) \,,\nonumber\\
(\pm)\,, \;\, {\rm if}\; m_{+}= (\mp)\, i\,m\,,  
% m^2&=& m_{+}\, m_{-}\,,\quad m_{+} = -m_{-}\,, \quad (p_{0})^2 =\vec{p}^{2}+ m^2\,.
\end{eqnarray}
is obviously equal to the particle and the antiparticle state at the same time, which is also true for 
any $\vec{p}$,  with the mass
\begin{eqnarray}
\label{Majoranamass}
-<\psi^{(6)}_{\pm \frac{1}{2},\tiny{majorana}} |\,\gamma^{0} \,(\stackrel{56}{(+)}\,m_{+} + 
\stackrel{56}{(-)}\,m_{-})|\, \psi^{(6)}_{\pm \frac{1}{2},majorana}> = m  
% m^2&=& m_{+}\, m_{-}\,,\quad m_{+} = -m_{-}\,, \quad (p_{0})^2 =\vec{p}^{2}+ m^2\,.
\end{eqnarray}

{\it Let us summarize}: The Majorana particle is, in this case of no charge, the particle and the 
antiparticle at the same time.

\section{Comments and conclusions}
\label{conclusions}

 This paper is written to stress some well known properties of quantum states. The properties of 
 particle  and antiparticle states are determined by the action: Starting with a well defined 
 action, which demonstrates in $d=(3+1)$ the charges and the masses,  so that the origin of both is well 
 understood, the discrete symmetry operators are (although  model dependent) well defined. 
 Changing the starting action into the effective one, the discrete symmetries will usually need  
 a redefinition.  The action as well the corresponding Weyl equations for massless particle (and 
 antiparticle) states contain  an even number of the Clifford operators -  
 $\gamma^a$ matrices, as it do also the Weyl equations for massless states in $d=(3+1)$. The 
 Clifford odd operators transform one Weyl representation into another one. 
 Analysing properties of states obtained from the solutions of the Weyl equation  with respect to 
 the Clifford odd operators makes sense, if the  starting action connects even and odd representations.
 The mass term, appearing with the Clifford odd operator in the Dirac equation, is the effective 
 one. The origin of the mass is  unknown. 
 
 We demonstrate in this paper on a toy model that whatever properties of particles  and antiparticles 
 are studied, the action is needed, the solutions of which are the states under consideration,
 if one wants to understand properties of states.

We discuss degrees of freedom of particles and antiparticles, starting with a well defined 
representation. We take the toy model in which the $M^{5+1}$ manifold 
breaks into $M^{3+1}\times$ an almost $S^2$ sphere due to the zweibein and the spin connection 
fields in $d=(5,6)$. We look for the solutions of the Weyl equations within one Weyl 
representation in $d=(5+1)$ and study the degrees of freedom and symmetries, which massless and 
massive particles and antiparticles manifest in $d=(3+1)$.

The toy model~\cite{DHN,DN012,HN,hn07} is chosen to make discussions as transparent as possible. 
The massless and mass protected spinors manifest the spin in $d=(5,6)$ as the (Kaluza-Klein) charge 
in $d=(3+1)$, while a spinor with the total angular momentum equal to $n+\frac{1}{2}$,  
$0< n \le l$, $l,0,1,2,\dots$, manifest $n+\frac{1}{2}$ as the  charge and carry the 
nonzero mass ($(m \rho_{0})^2=l(l+1)$, $\rho_{0}$ is the radius of an almost $S^2$) with respect 
to $d=(3+1)$. The spinors states differ in masses and charges. We also treat the case when the spin 
connections in with the indices $(5,6)$, and correspondingly scalars with respect to $d=(3+1)$, 
gain non zero vacuum expectation values. In this case there are no massless particles any longer, 
as well as no conserved charges  massive chargeless particles in $d=(3+1)$ behave as 
the Majorana particles.

We use the technique~\cite{norma92,norma93,hno2,hno6,norma94,norma95,pikanorma} 
for representing spinors, which makes the illustration transparent.
We use the concept of the Dirac sea to treat the second quantized picture, what enables a nice 
physical understanding. The discrete symmetry operators for the first and the second quantized 
picture are taken from the ref.~\cite{HNds,NBled2013}.

We conclude that in all the studied cases, in the case of the massless (with $U(1)$) charged states 
in the case of the massive (with $U(1)$) charged states and in the massive chargeless case, 
there are two positive and two negative energy solutions of the equations of motion. All the states,
the particle states, the negative energy states in the Dirac sea, as well as the antiparticle
states, which are the holes in the Dirac sea,  solve the equations of motion. Either the negative 
energy states or the positive energy states are obtainable also directly  from the particle states 
by the application of the discrete symmetry operators~\cite{NBled2013,HNds,TDN} 
${\cal C}_{{\cal N}}\cdot $ ${\cal P}^{(d-1)}_{{\cal N}}$ and 
$\mathbb{C}_{{\cal N}}\cdot {\cal P}^{(d-1)}_{{\cal N}}$, 
respectively.

As long as the charges are conserved quantities, the antiparticle states distinguish from the particle 
states in charges and masses, as expected. The antiparticle states are of the opposite handedness 
in $d=(3+1)$ as the particle states, they both are of the same handedness in $d=(5+1)$. 

In the case that the vacuum expectation values of the scalar fields, the analogue of the higgs in the 
{\it standard model} carrying in the toy model case the hyper charge only (it is the $U(1)$ charge 
of the integer value), causing that no charge is conserved and that massless solutions become massive,
the antiparticle states coincide with the particle states, representing the Majorana particles.

 If one assumes $d=(7+1)$ instead of $d=(5+1)$ and lets all the scalar (with the indices $(5,6,7,8)$) spin 
 connection fields to  gain nonzero vacuum expectation values, it is still true that there are two positive 
 (with spin $\pm \frac{1}{2}$) and two negative (again with spin $\pm \frac{1}{2}$) solutions of the 
 Weyl equations of motion, but in this case the 
antiparticle states distinguish from the particle states~\footnote{One must pay attention that in 
even dimensional cases with $d=4n$ the operator $\mathbb{C}_{(\cal N)}$   has 
an even number of $\gamma^a$'s.} rather then $\mathbb{C}_{(\cal N)}$ $\times $ $ {\cal P}^{(d-1)}_{{\cal N}}$,
which is the case in $d=2(2n+1)$ and is correspondingly a good discrete symmetry of the system. 

The higher is the dimension the more charges are available. Taking  $d=(13+1)$, the action of 
Eq.~(\ref{action}) has in the starting Weyl representation all the particle and antiparticle states as 
required by the {\it standard model}, with the right handed neutrinos added. The right handed quarks 
and leptons, weak chargeless,  carry the additional $SU(2)$ 
charge which after the break of the starting symmetries manifest the hyper charge, while the 
left handed ones are weak charged and carry no $SU(2)$ charge of the second kind 
and consequently no hyper charge. The quarks distinguish from the leptons  besides in the colour charge 
also in the "fermion" number, which is $\frac{1}{6}$ for quarks and $-\frac{1}{2}$ for leptons, 
while the antiquark and the antilepton states, appearing in the same Weyl representation,  carry the 
opposite  weak,  colour, 
additional $SU(2)$ charge and the  "fermion" quantum number.  Correspondingly is the "fermion" charge
equal to zero for particles and antiparticles separately. 

The discrete symmetry operators  of Eq.~(\ref{CPTN},\ref{CemptPTN}) have the desired properties 
also in the case of $d=(13+1)$  (as they are in all even dimensional spaces). All the 
antiparticle states are reachable from the particle ones by the application of the operator 
$\mathbb{C}_{(\cal N)}$ $\times $ $ {\cal P}^{(d-1)}_{{\cal N}}$. (The operator of "emptying" is, 
although not easy to imagine, since it operate among two completely different Fock spaces, 
very useful). All the negative energy states 
are reachable by the application of ${\cal C}_{(\cal N)}$ $\times $ $ {\cal P}^{(d-1)}_{{\cal N}}$.
$\mathbb{C}_{(\cal N)}$ $\times $ $ {\cal P}^{(d-1)}_{{\cal N}}$ is the conserved quantity, unless 
the families are taken into account and the scalar fields gain nonzero vacuum expectation values, 
offering $\frac{(n-1)(n-2)}{2}$ complex phases. The scalar fields with the space index $(7,8)$ carry the 
$SU(2)$ weak charge ($\pm \frac{1}{2}$) and the hyper charge ($\mp \frac{1}{2}$)  like in the {\it 
standard model}. 

Although in $d=(3+1)$ there exist the left and the right handed solutions (an example is presented 
Eq.~(\ref{weylgen})), one can never come from one to another solution by the application of an 
Clifford odd operator. The change of the handedness in $d=(3+1)$ is always accompanied by the change 
of a spin in higher dimensions (representing a charge in $d=(3+1)$.

We conclude from this discussions that analysing the Dirac states in $d=(3+1)$ without having a 
model behind, telling where do the effective masses and charges originate, might be misleading. 
We hope that our discussions in this paper will help to clarify the matter.

\appendix*
\section{The technique for representing spinors~\cite{HNds,NBled2013,norma}, a shortened version 
of the one presented in~\cite{NBled2013,norma}}
\label{technique}

The technique~\cite{norma92,norma93,hno2,hno6,norma94,norma95,pikanorma,HNds,norma} can be used to 
construct a spinor basis for any dimension $d$
and any signature in an easy and transparent way. Equipped with the graphic presentation of basic states,  
the technique offers an elegant way to see all the quantum numbers of states with respect to the  
Lorentz groups, as well as transformation properties of the states under any Clifford algebra object.

The objects $\gamma^a$ %and $\tilde{\gamma}^a$ 
have properties %~(\ref{snmb:tildegclifford}),
%
%\begin{eqnarray}
%\label{gammatildegamma}&& 
$\{ \gamma^a, \gamma^b\}_{+} = 2\eta^{ab}\,I, $
%
%\quad\quad    
%\{ \tilde{\gamma}^a, \tilde{\gamma}^b\}_{+}= 2\eta^{ab}\,, \quad,\quad
%\{ \gamma^a, \tilde{\gamma}^b\}_{+} = 0\,,
%\end{eqnarray}
%
for any $d$, even or odd.  $I$ is the unit element in the Clifford algebra.

The Clifford algebra objects $S^{ab}$  %and $\tilde{S}^{ab}$ 
close the algebra of the Lorentz group 
%
%\begin{eqnarray}
%\label{sabtildesab}
$S^{ab}:  = (i/4) (\gamma^a \gamma^b - \gamma^b \gamma^a)\,$, %\nonumber\\
%\tilde{S}^{ab}: &=& (i/4) (\tilde{\gamma}^a \tilde{\gamma}^b 
%- \tilde{\gamma}^b \tilde{\gamma}^a)\,,\nonumber\\
 %\{S^{ab}, \tilde{S}^{cd}\}_{-}&=& 0\,,\nonumber\\
$\{S^{ab},S^{cd}\}_{-}  = i(\eta^{ad} S^{bc} + \eta^{bc} S^{ad} - \eta^{ac} S^{bd} - \eta^{bd} S^{ac})\,$.
%\nonumber\\
%\{\tilde{S}^{ab},\tilde{S}^{cd}\}_{-} &=& i(\eta^{ad} \tilde{S}^{bc} + \eta^{bc} \tilde{S}^{ad} 
%- \eta^{ac} \tilde{S}^{bd} - \eta^{bd} \tilde{S}^{ac})\,.
%\end{eqnarray}
%
The ``Hermiticity'' property for $\gamma^a$'s:   
$\,\gamma^{a\dagger} = \eta^{aa} \gamma^a\,$ 
%\quad \quad \tilde{\gamma}^{a\dagger} = \eta^{aa} \tilde{\gamma}^a\,,$
is assumed  in order that $\gamma^a$ are 
%
%are compatible with (\ref{gammatildegamma}) and 
formally unitary, 
i.e. $\gamma^{a \,\dagger} \,\gamma^a=I$. %and $\tilde{\gamma}^{a\,\dagger} \tilde{\gamma}^a=I$.
%
%One finds %from~(\ref{cliffher}) 
%that $(S^{ab})^{\dagger} = \eta^{aa} \eta^{bb}S^{ab}$. 

The Cartan subalgebra of the algebra %two groups, which  form  equivalent representations  
is chosen in even dimensional spaces  as follows: 
%
%\begin{eqnarray}
$\,S^{03}, S^{12}, S^{56}, \cdots, S^{d-1\; d}, \quad {\rm if } \quad d  = 2n \ge 4$.
%\nonumber\\
%S^{03}, S^{12}, \cdots, S^{d-2 \;d-1}, \quad {\rm if } \quad d &=& (2n +1) >4\,.
%\nonumber\\
%\tilde{S}^{03}, \tilde{S}^{12}, \tilde{S}^{56}, \cdots, \tilde{S}^{d-1\; d}, 
%\quad {\rm if } \quad d &=& 2n\ge 4\,,
%\nonumber\\
%\tilde{S}^{03}, \tilde{S}^{12}, \cdots, \tilde{S}^{d-2 \;d-1}, 
%\quad {\rm if } \quad d &=& (2n +1) >4\,.
%\label{choicecartan}
%\end{eqnarray}
%

The choice for  the Cartan subalgebra in $d > 4$ is straightforward.
It is  useful  to define one of the Casimir operators of the Lorentz group -  
the  handedness $\Gamma$ ($\{\Gamma, S^{ab}\}_- =0$) in any $d$, for  even dimensional spaces  it follows: 
%
%\begin{eqnarray}
$\Gamma^{(d)} :=(i)^{d/2}\; \;\;\;\;\;\prod_a \quad (\sqrt{\eta^{aa}} \gamma^a), \quad {\rm if } \quad d = 2n\,$. 
%\nonumber\\
%\Gamma^{(d)} :&=& (i)^{(d-1)/2}\; \prod_a \quad (\sqrt{\eta^{aa}} \gamma^a), \quad {\rm if } \quad d = 2n +1\,.
%\label{hand}
%\end{eqnarray}
%
The product of $\gamma^a$'s in the ascending order with respect to 
the index $a$: $\gamma^0 \gamma^1\cdots \gamma^d$ is understood. 
It follows %from~(\ref{cliffher})
for any choice of the signature $\eta^{aa}$ that
$\Gamma^{\dagger}= \Gamma,\;
\Gamma^2 = I.$
For $d$ even the handedness  anticommutes with the Clifford algebra objects 
$\gamma^a$ ($\{\gamma^a, \Gamma \}_+ = 0$).
%
%(while for $d$ odd it commutes with  
%$\gamma^a$ ($\{\gamma^a, \Gamma \}_- = 0$)). 

To make the technique simple  the graphic presentation is introduced
\begin{eqnarray}
\stackrel{ab}{(k)}:&=& 
\frac{1}{2}(\gamma^a + \frac{\eta^{aa}}{ik} \gamma^b)\,,\quad \quad
\stackrel{ab}{[k]}:=
\frac{1}{2}(1+ \frac{i}{k} \gamma^a \gamma^b)\,,
%\nonumber\\
%\stackrel{+}{\circ}:&=& \frac{1}{2} (1+\Gamma)\,,\quad \quad
%\stackrel{-}{\bullet}:= \frac{1}{2}(1-\Gamma),
\label{signature}
\end{eqnarray}
where $k^2 = \eta^{aa} \eta^{bb}$.
One can easily check by taking into account the  Clifford algebra relation 
%(\ref{gammatildegamma}) 
and the definition of $S^{ab}$ %and $\tilde{S}^{ab}$ 
%(\ref{sabtildesab})
that if one multiplies from the left hand side by $S^{ab}$ %or $\tilde{S}^{ab}$ 
the Clifford algebra objects $\stackrel{ab}{(k)}$ and $\stackrel{ab}{[k]}$, it follows that
\begin{eqnarray}
        S^{ab}\, \stackrel{ab}{(k)}= \frac{1}{2}\,k\, \stackrel{ab}{(k)}\,,\quad \quad 
        S^{ab}\, \stackrel{ab}{[k]}= \frac{1}{2}\,k \,\stackrel{ab}{[k]}\,,%\nonumber\\
%\tilde{S}^{ab}\, \stackrel{ab}{(k)}= \frac{1}{2}\,k \,\stackrel{ab}{(k)}\,,\quad \quad 
%\tilde{S}^{ab}\, \stackrel{ab}{[k]}=-\frac{1}{2}\,k \,\stackrel{ab}{[k]}\,,
\label{grapheigen}
\end{eqnarray}
which means that we get the same objects back multiplied by the constant $\frac{1}{2}k$.
This also means that when 
$\stackrel{ab}{(k)}$ and $\stackrel{ab}{[k]}$ act from the left hand side on  a
vacuum state $|\psi_0\rangle$ the obtained states are the eigenvectors of $S^{ab}$.
One can further recognize %~(\ref{snmb:graphgammaaction},\ref{snmb:gammatilde}) 
that $\gamma^a$ transform  $\stackrel{ab}{(k)}$ into  $\stackrel{ab}{[-k]}$, never to $\stackrel{ab}{[k]}$:  
\begin{eqnarray} 
\gamma^a \stackrel{ab}{(k)}&=& \eta^{aa}\stackrel{ab}{[-k]},\; 
\gamma^b \stackrel{ab}{(k)}= -ik \stackrel{ab}{[-k]}, \; 
\gamma^a \stackrel{ab}{[k]}= \stackrel{ab}{(-k)},\; 
\gamma^b \stackrel{ab}{[k]}= -ik \eta^{aa} \stackrel{ab}{(-k)}\,.
\label{snmb:gammatildegamma}
\end{eqnarray}
%
%From~(\ref{snmb:gammatildegamma}) it follows
%%
%\begin{eqnarray}
%\label{stildestrans}
%S^{ac}\stackrel{ab}{(k)}\stackrel{cd}{(k)} &=& -\frac{i}{2} \eta^{aa} \eta^{cc} 
%%\stackrel{ab}{[-k]}\stackrel{cd}{[-k]}\,,\quad\quad
%%\tilde{S}^{ac}\stackrel{ab}{(k)}\stackrel{cd}{(k)} = \frac{i}{2} \eta^{aa} \eta^{cc} 
%%\stackrel{ab}{[k]}\stackrel{cd}{[k]}\,,\,
%%\nonumber\\
%S^{ac}\stackrel{ab}{[k]}\stackrel{cd}{[k]} = \frac{i}{2}  
%%\stackrel{ab}{(-k)}\stackrel{cd}{(-k)}\,,%\quad\quad
%%\tilde{S}^{ac}\stackrel{ab}{[k]}\stackrel{cd}{[k]} = -\frac{i}{2}  
%%\stackrel{ab}{(k)}\stackrel{cd}{(k)}\,,\,
%\nonumber\\
%S^{ac}\stackrel{ab}{(k)}\stackrel{cd}{[k]}  &=& -\frac{i}{2} \eta^{aa}  
%\stackrel{ab}{[-k]}\stackrel{cd}{(-k)}\,,\quad\quad
%%\tilde{S}^{ac}\stackrel{ab}{(k)}\stackrel{cd}{[k]} = -\frac{i}{2} \eta^{aa}  
%%\stackrel{ab}{[k]}\stackrel{cd}{(k)}\,,\,\nonumber\\
%S^{ac}\stackrel{ab}{[k]}\stackrel{cd}{(k)} = \frac{i}{2} \eta^{cc}  
%\stackrel{ab}{(-k)}\stackrel{cd}{[-k]}\,. %\,\quad\quad
%%\tilde{S}^{ac}\stackrel{ab}{[k]}\stackrel{cd}{(k)} = \frac{i}{2} \eta^{cc}  
%%\stackrel{ab}{(k)}\stackrel{cd}{[k]}\,. 
%\end{eqnarray}
%% 

Let us deduce some useful relations
\begin{eqnarray}
\stackrel{ab}{(k)}\stackrel{ab}{(k)}& =& 0\,, \quad \quad \stackrel{ab}{(k)}\stackrel{ab}{(-k)}
= \eta^{aa}  \stackrel{ab}{[k]}\,, \quad  %\stackrel{ab}{(-k)}\stackrel{ab}{(k)}=
%\eta^{aa}   \stackrel{ab}{[-k]}\,,\quad
%\stackrel{ab}{(-k)} \stackrel{ab}{(-k)} = 0\,, \nonumber\\
 \stackrel{ab}{[k]}\stackrel{ab}{[k]} = \stackrel{ab}{[k]}\,, \quad \quad
\stackrel{ab}{[k]}\stackrel{ab}{[-k]}= 0\,,% \;\;\quad \quad  \quad \stackrel{ab}{[-k]}\stackrel{ab}{[k]}=0\,,
 %\;\;\quad \quad \quad \quad \stackrel{ab}{[-k]}\stackrel{ab}{[-k]} = \stackrel{ab}{[-k]}\,,
 \nonumber\\
 \stackrel{ab}{(k)}\stackrel{ab}{[k]}& =& 0\,,\quad \quad \quad \stackrel{ab}{[k]}\stackrel{ab}{(k)}
=  \stackrel{ab}{(k)}\,, \quad \quad %\quad \stackrel{ab}{(-k)}\stackrel{ab}{[k]}=
% \stackrel{ab}{(-k)}\,,\quad \quad \quad 
%\stackrel{ab}{(-k)}\stackrel{ab}{[-k]} = 0\,,
%\nonumber\\
 \stackrel{ab}{(k)}\stackrel{ab}{[-k]}=  \stackrel{ab}{(k)}\,,
\quad \quad \stackrel{ab}{[k]}\stackrel{ab}{(-k)} =0\,. %  \quad \quad 
%\quad \stackrel{ab}{[-k]}\stackrel{ab}{(k)}= 0\,, \quad \quad \quad \quad
%\stackrel{ab}{[-k]}\stackrel{ab}{(-k)} = \stackrel{ab}{(-k)}.
\label{graphbinoms}
\end{eqnarray}
%
%We recognize in  the first equation of the first line and the third equation of the first line
%the demonstration of the nilpotent and the projector character of the Clifford algebra objects 
%$\stackrel{ab}{(k)}$ and $\stackrel{ab}{[k]}$, respectively. 
%%
%Recognizing that
%%
%%\begin{eqnarray}
%$\stackrel{ab}{(k)}^{\dagger}=\eta^{aa}\stackrel{ab}{(-k)}\,$,
%$\stackrel{ab}{[k]}^{\dagger}= \stackrel{ab}{[k]}\,$,
%%\label{graphherstr}
%%\end{eqnarray}
%%
% a vacuum state $|\psi_0>$ can be defined so that it follows
%%
%\begin{eqnarray}
%< \;\stackrel{ab}{(k)}^{\dagger}
% \stackrel{ab}{(k)}\; > = 1\,,\quad \quad < \;\stackrel{ab}{[k]}^{\dagger}
% \stackrel{ab}{[k]}\; > = 1\,.
%\label{graphherscal}
%\end{eqnarray}
%%

Taking into account the above equations it is easy to find a Weyl spinor irreducible representation
for $d$-dimensional space. %, with $d$ even or odd. 

For $d$ even we simply make a starting state as a product of $d/2$, let us say, only nilpotents 
$\stackrel{ab}{(k)}$, one for each $S^{ab}$ of the Cartan subalgebra  elements, 
%(\ref{choicecartan}),  
applying it on an (unimportant) vacuum state. 
%For $d$ odd the basic states are products
%of $(d-1)/2$ nilpotents and a factor $(1\pm \Gamma)$.  
Then the generators $S^{ab}$, which do not belong 
to the Cartan subalgebra, being applied on the starting state from the left, 
 generate all the members of one
Weyl spinor.  
\begin{eqnarray}
\stackrel{0d}{(k_{0d})} \stackrel{12}{(k_{12})} \stackrel{35}{(k_{35})}\cdots \stackrel{d-1\;d-2}{(k_{d-1\;d-2})}
\psi_0 \nonumber\\
\stackrel{0d}{[-k_{0d}]} \stackrel{12}{[-k_{12}]} \stackrel{35}{(k_{35})}\cdots \stackrel{d-1\;d-2}{(k_{d-1\;d-2})}
\psi_0 \nonumber\\
%
%\stackrel{0d}{[-k_{0d}]} \stackrel{12}{(k_{12})} \stackrel{35}{[-k_{35}]}\cdots \stackrel{d-1\;d-2}{(k_{d-1\;d-2})}
%\psi_0 \nonumber\\
\vdots \nonumber\\
%\stackrel{0d}{[-k_{0d}]} \stackrel{12}{(k_{12})} \stackrel{35}{(k_{35})}\cdots \stackrel{d-1\;d-2}{[-k_{d-1\;d-2}]}
%\psi_0 \nonumber\\
\stackrel{0d}{(k_{0d})} \stackrel{12}{[-k_{12}]} \stackrel{35}{[-k_{35}]}\cdots \stackrel{d-1\;d-2}{(k_{d-1\;d-2})}
\psi_0 \nonumber\\
\vdots 
\label{graphicd}
\end{eqnarray}
All the states have the handedness $\Gamma $, since $\{ \Gamma, S^{ab}\}_{-} = 0$. 
States, belonging to one multiplet  with respect to the group $SO(q,d-q)$, that is to one
irreducible representation of spinors (one Weyl spinor), can have any phase. We made a choice
of the simplest one, taking all  phases equal to one.

%The above graphic representation demonstrates that for $d$ even 
%all the states of one irreducible Weyl representation of a definite handedness follow from a starting state, 
%which is, for example, a product of nilpotents $\stackrel{ab}{(k_{ab})}$, by transforming all possible pairs
%of $\stackrel{ab}{(k_{ab})} \stackrel{mn}{(k_{mn})}$ into $\stackrel{ab}{[-k_{ab}]} \stackrel{mn}{[-k_{mn}]}$.
%There are $S^{am}, S^{an}, S^{bm}, S^{bn}$, which do this.
%The procedure gives $2^{(d/2-1)}$ states. A Clifford algebra object $\gamma^a$ being applied from the left hand side,
%transforms  a 
%Weyl spinor of one handedness into a Weyl spinor of the opposite handedness. Both Weyl spinors form a Dirac 
%spinor. 

\section*{Acknowledgments} One of the authors acknowledges funding of the Slovenian Research Agency.

\end{document}